\documentclass[twocolumn,twoside]{article}

\usepackage{a4}
\usepackage{amssymb}
\usepackage{amsmath}
\usepackage[numbers,sort&compress]{natbib}
\usepackage{graphicx}

\begin{document}


\title{{\bf Radio pulsars: the search for truth}}

\date{{\normalsize\textit{P N Lebedev Physical Institute, 
Leninskii prosp. 53, Moscow, 119991,
\\
Moscow Institute of Physics and Technology, Institutsky per, 9, 
Dolgoprudny, Moscow region,  141700, 
\\
Department of Astrophysical Sciences, 
Princeton University,
4 Ivy Lane, Peyton Hall, Princeton, NJ 08544
}}\\[1ex]
{\small \textit{Usp.\ Fiz.\ Nauk} \textbf{183}, 179--194 (2013)
[in Russian]\\
English translation: \textit{Physics -- Uspekhi}, \textbf{56}, 164--179 (2013)}
\\{\small Translated by G Pontecorvo; edited by A Semikhatov}
}

\author{V S Beskin, Ya N Istomin, A A Philippov}

\maketitle

{\bf \underline{Abstract.} {\small It was as early as the 1980s that A V Gurevich and his group 
proposed a theory to explain the magnetosphere of radio pulsars and the mechanism by which they 
produce coherent radio emission. The theory has been sharply criticized and is currently rarely 
mentioned when discussing the observational properties of radio pulsars, even though all the 
criticisms were in their time disproven in a most thorough and detailed manner. Recent results 
show even more conclusively that the theory has no internal inconsistencies. New observational 
data also demonstrate the validity of the basic conclusions of the theory. Based on the latest 
results on the effects of wave propagation in the magnetosphere of a neuron star, we show that 
the developed theory does indeed allow quantitative predictions of the evolution of neutron stars 
and the properties of the observed radio emission.}} 

\setcounter{secnumdepth}{3}
\setcounter{tocdepth}{2}

\tableofcontents

\section{Introduction} 

Thirty years have passed since the group led by A V Gurevich at the Theoretical Physics 
Department of the Lebedev Physical Institute, Russian Academy of Sciences, published its 
first article [1] on the theory of the magnetosphere of radio pulsars, 25 years since the 
publication of their article [2] dealing with the development of the theory of radio 
emission of pulsars, and, finally, 20 years since the publication of mongraph [3], in 
which a consistent theory was formulated of the principal processes responsible for 
the observed activity of radio pulsars. In hindsight, we would like to make some comments, 
which, we hope, can be useful at the present stage of development of the theory.

Currently, already more than 40 years since radio pulsars were discovered [4] in 1967, 
our understanding of these objects remains somewhat ambiguous. On the one hand, significant 
progress had been achieved already during the first decade after the discovery of radio 
pulsars, when the theory of the magnetosphere of radio pulsars and the theory of their 
radio emission were being actively developed by leading scientists throughout the world: 
Ginzburg [5], Zheleznyakov [6], Kadomtsev [7], Sagdeev [8], and Lominadze [9] in the USSR, 
and Goldreich [10], Coppi [11], Melrose [12], Mestel [13], and Ruderman [14] in other countries. 
That was a period of 'storm and stress' ('Sturm und Drang'), which permitted anticipating the 
main properties of radio pulsars (see, e.g., monographs [15, 16]); the extremely stable sequence 
of radio pulses is due to the rotation of a neutron star, the kinetic energy of rotation is 
the source of energy, while the mechanism reponsible for the slowing down of rotation is of 
an electromagnetic nature.

Total success was not achieved, however, in spite of a number of tactical gains (for instance, 
the key role of the secondary electron-positron plasma filling the magnetosphere of a neutron 
star was fully appreciated [14,17], as also was the importance of the current mechanism of 
energy losses, i.e., of energy losses due to longitudinal currents flowing in the magnetosphere 
of a pulsar [10]). In particular, it is only in recent years that processes occurring in the 
magnetosphere of an oblique rotator have started to be actively discussed [18-21]; previously, 
most work was devoted to the axisymmetric magnetosphere [22-29]. No common standpoint regarding 
the mechanism of coherent radio emission exists yet [30-32].

By the late 1980s, research activities on the theory of the pulsar magnetosphere and the theory 
of radio emission were drastically reduced. Two or three serious publications per year (!) 
certainly made no real difference. Actually, a period of deep stagnation set in. On the one hand, 
most theorists were not able to propose a model that could provide readily testable predictions, 
while, on the other hand, the existence of a simple hollow-cone geometric model [33] (in which 
the directivity pattern of the radio emission repeats the density profile of the secondary 
electron-positron plasma outflowing along the open magnetic field lines) permitted interpreting 
observational data without turning to theory. As a result, the connection between theory and 
observations was practically lost.

Perhaps, precisely the atmosphere of general failure (although there most likely existed purely 
opportunistic reasons, also) resulted in a series of studies, performed in the 1980s by the group 
led by Gurevich [1, 2, 34, 35], in which the authors succeeded in formulating a consistent theory 
of the magnetosphere and of the radio emission of pulsars, which was met, mildly speaking, without 
friendliness. Here, we only present a few quotations from articles and reviews on book [3] that 
was the conclusion of a decade of work{\footnote{Incidentally, the fact that critical reviews 
continued to appear until quite recently [36, 37] rather serves as an argument in favor of our 
conclusions.}}. "We conclude that their computation of the dielectric tensor of a plasma in a 
strong magnetic field is wrong" [38]. "It has been claimed that this instability is spurious" 
[39]. "This theory is known to be incorrect. It contains several fatal flows" [40].

Regretfully, such peremptory criticism made any serious discussion of the results of our work 
simply impossible, although practically all the main critical pronouncements were given detailed 
explanations [30, 41, 42], demonstrating their judgements to be unjustified. Therefore, as before, 
we believe in the validity of our conclusions, which can be formulated as follows.

\vspace*{.3cm}

I. Theory of the magnetosphere of radio pulsars.
\begin{enumerate}
\item
The plasma filling the pulsar magnetosphere totally screens the magnetodipole radiation 
of a neutron star. As a result, all the energy losses must be due to longitudinal currents 
circulating within the pulsar magnetosphere (and closing up on the surface of the neutron 
star).
\item
For a local Goldreich current (see Section 2.2), current losses should be significantly 
smaller for an orthogonal rotator than for the axisymmetric one, which follows from 
the expression we found for current losses for an arbitrary inclination angle of the magnetic 
dipole axis to the axis of rotation.
\item
The inclination angle of the magnetic dipole axis to the axis of rotation should increase 
with time, unlike the magnetodipole losses.
\item
Quite a small longitudinal current is realized in the pulsar magnetosphere, which results 
in the inevitable appearance outside the light cylinder of a region where the electric field 
is greater than the magnetic field. Inside this region, effective acceleration of the plasma 
flowing out of the pulsar magnetosphere becomes possible.
\end{enumerate} 

II. Theory of radio emission.
\begin{enumerate}
\item
The dielectric permittivity tensor of plasma in an inhomogeneous medium has been found, 
in particular, for the relativistic plasma moving along curved magnetic field lines. 
This tensor permits correctly describing the interaction of particles with waves propagating 
in an inhomogeneous plasma.
\item
The analysis of such a tensor reveals the instability of 'curvature-plasma' waves, which can 
serve as the base instability for the maser mechanism providing coherence of the observed 
emission.
\item
By taking the nonlinear interaction of waves into account, the excitation level has been 
determined for transverse waves capable to escape from the magnetosphere of a neutron star, 
and, thus, the intensity of the radio emission of pulsars has been established.
\item
Based on a consistent analysis of wave propagation in the magnetosphere of a pulsar, which, 
for instance, takes the refraction of an ordinary wave into account, quantitative predictions 
have been made concerning the main observational characteristics of pulsars (the frequency 
dependence of the width of mean profiles, the statistics of pulsars with single or double 
profiles, etc.).
\end{enumerate}
The purpose of this article is to show that at present, sufficient material has been 
accumulated to assert with confidence that the principal theoretical points of our theory 
not only have not become obsolete (which could well have happened owing to the impetuous 
development of numerical methods) but also can be a basis for understanding the phenomenon 
of radio pulsars. Moreover, we show that the recently obtained observational data confirm 
the validity of the main theoretical conclusions formulated over 20 years ago.

\section {Pulsar magnetosphere}

\subsection {Screening of the magnetodipole radiation}

A uniformly magnetized sphere rotating in the vacuum is known to lose rotational energy 
owing to magnetodipole losses [43]:
\begin{equation}
W_{\rm tot} = -J_{\rm r}\Omega\dot\Omega =
\frac{1}{6}\frac{B_0^2\Omega^4R^6}{c^3}\sin^2\chi.
\label{wmd}
\end{equation}
Here, $R$ is the radius of the sphere, $B_0$ is the magnetic field at the magnetic pole, 
$J_{\rm r} \sim MR^2$ is the moment of inertia, $\chi$  is the inclination angle of the 
magnetic axis to the rotation axis, and $\Omega = 2\pi/P$ is the angular velocity 
of rotation. This mechanism is quite universal, and hence it would be natural to assume 
expression (1) for magnetodipole losses to also hold in the case of a magnetosphere filled 
with secondary electron-positron plasma. Therefore, estimate (1) still expresses the common 
view of the rotation deceleration mechanism of radio pulsars.

However, this conclusion, seemingly evident at first glance, turned out to have no foundation. 
To be more precise, we showed that in the framework of a force-free approximation (a secondary 
plasma, whose energy density is negligible compared to the energy density of the magnetic field, 
fully screens the longitudinal electric field) and in the case of a zero longitudinal (parallel 
to the magnetic field) electric current, the flux of the Poynting vector through the surface of 
a light cylinder, $R_{\rm L} = c/\Omega$, vanishes [1]. From a mathematical standpoint, this is 
because the toroidal component of the magnetic field on the surface of the light cylinder must 
vanish (actually, this conclusion was obtained in 1975 in Ref. [44]): 
\begin{equation}
B_{\varphi}(R_{\rm L}) = 0.
\label{Bphi}
\end{equation}
From a physical standpoint, this means that the plasma filling the pulsar magnetosphere 
completely screens the magnetodipole radiation of the neutron star. In other words, in 
the case of a zero longitudinal current, the magnetospheric plasma emission is in a phase 
precisely opposite to the phase of the pulsar magnetodipole radiation. Consequently, all 
the energy losses should be due to longitudinal currents circulating inside the magnetosphere 
of the neutron star and closing up on its surface. These energy losses can be determined by 
the formula $W_{\rm tot} = -{\bf \Omega} {\bf K}$, where
\begin{equation}
{\bf K} =
\frac{1}{c}\int[{\bf r}\times[{\bf J}_{\rm s}\times{\bf B}]]{\rm d}S
\label{e3}
\end{equation}
is the decelerating moment of the Ampere force due to surface currents $J_{\rm s}$. 
We recall that it is possible to obtain an analytic solution for an oblique rotator because 
in the case of a zero longitudinal current, the equation describing the magnetosphere of a 
neutron star is linear; it is also extremely important here that the boundary condition at 
infinity (along the rotation axis) was used.

It follows that the energy losses can be written as [1]
\begin{equation}
W_{\rm tot} = \frac{f_*^2}{4} \, \frac{B_0^2\Omega^{4}R^6}{c^3}i_0\cos\chi,
\label{e15}
\end{equation}
where $i_0 = j_{\parallel}/j_{\rm GJ}$ is the dimensionless longitudinal current normalized 
to the so-called Goldreich current (or the Golreich-Julian current),
\begin{equation}
j_{\rm GJ} = \frac{\Omega B}{2 \pi},
\label{jgj}
\end{equation}
where $f_{\ast} = 1.59$--$1.96$ is a coefficient that depends weakly on the inclination angle 
$\chi$.

We note that the conclusion concerning the complete screening of the magnetodipole radiation 
was, naturally, not obvious. Therefore, not surprisingly, it remains far from being adopted 
by everyone. It is interesting that now, after 30 years have passed, we have been surprised 
to learn from many participants in those discussions that the main criticism seems to have 
concerned our alleged claim that pulsars lose no rotation energy at all. But we never made 
any such statement and could not have done so. The main conclusion in Ref. [1] is formula 
(4) for current energy losses, which clearly points to the deceleration mechanism.

At present, screening of the magnetodipole radiation of a neutron star can be confidently 
said to indeed take place. First of all, in 1999, the group of L Mestel [45] performed studies 
equivalent to those presented in Ref. [1] and fully confirmed our conclusions: in the case 
of a zero longitudinal current, the energy losses of an oblique rotator are equal to zero. 
Fig. 1 shows the structure of the magnetic field of an orthogonal rotator in the equatorial 
plane, obtained in Ref. [45] (the corresponding cross section remained in the draft copies 
of Ref. [1]). It is clearly seen that the magnetic field lines indeed approach the light 
cylinder at a right angle.

However, the most important recent result is that magnetodipole losses are also absent in 
the solution for the magnetosphere of an oblique rotator constructed by Spitkovsky on the 
basis of numerical simulation [20]. First of all, this follows from the assertion concerning 
the split-monopole asymptotic form of the solution obtained, which is close to the model of 
radial magnetized wind [18]. Such flows only involve stationary magnetized wind (independent 
of time outside a thin current sheet), in which, however, the energy flux is related to the 
flux of the Poynting vector. But the main point is that the absence of magnetodipole losses 
is also confirmed by a straightforward analysis of the structure of electromagnetic fields in 
Ref. [20].

\begin{figure}
\begin{center}
\includegraphics[width=6cm]{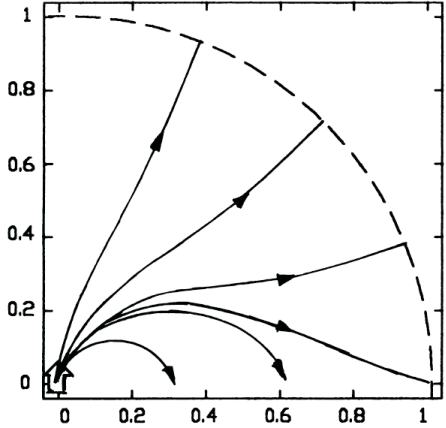}
\end{center}
\caption{\small {Structure of magnetic field lines of an orthogonal rotator in the equatorial 
plane [45]. The toroidal magnetic field is zero on the light cylinder.}
   }
\label{fig1_01}
\end{figure}

Indeed, in the case of a vacuum rotator, for any inclination angle  $\chi \neq 0^{\circ}$, 
a variable-in-time component of the magnetic field must be present on the rotation axis, 
with the amplitude
\begin{equation}
B_{\perp} = \frac{|{\ddot  {\bf m}}|}{c^2 r},
\end{equation}
where ${\bf m}$ is the magnetic moment of the star  (with $|{\bf m}| = B_0 R^{3}/2$), 
${\ddot {\bf m}}$ is its second time derivative, and 
\mbox{$|{\ddot {\bf m}}| = |{\bf m}|\Omega^2 \sin\chi$.}
However, as can be seen from Fig. 2, the variable component of the magnetic field in the 
Spitkovsky solution decreases much more rapidly, like $1/r^3$. In our opinion, the absence 
of variable fields decreasing as  $1/r$ in the numerical solution for an oblique rotator 
is the most convincing proof of the total screening of magnetodipole radiation in the case 
of a magnetosphere completely filled with plasma.

\subsection {Current losses}

One more important consequence of the theory of current losses is that for a local longitudinal 
Goldreich current, the rotational energy losses should decrease as the inclination angle
$\chi$ increases [1, 3]. The point is that, besides the factor $\cos \chi$ related to 
the scalar product $W_{\rm tot} = -{\bf \Omega} \cdot {\bf K}$, a significant dependence of 
the current losses $W_{\rm tot}$ in Eqn (4) on the inclination angle is also involved in 
the quantity $i_0$. 

\begin{figure}
\begin{center}
\includegraphics[width=7cm]{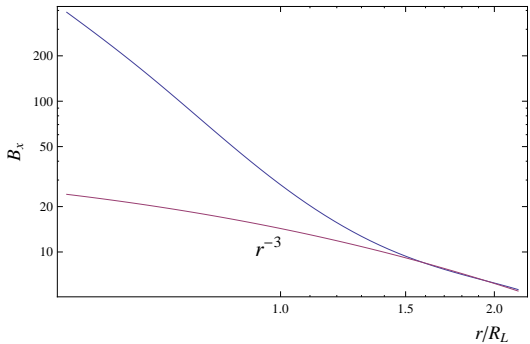}
\end{center}
\caption{\small{Dependence of the $B_{x}$ component of the magnetic field on the rotation 
axis on the distance $r$ from the neutron star in numerical simulation [20]. The lower 
curve shows the asymptotic behaveour $B \propto r^{-3}$. The inclination angle is 
$\chi = 60^{\circ}$.}
}
\label{fig1_2}
\end{figure}

Indeed, in the definition of the dimensionless current $i_0 = j_{\parallel}/j_{\rm GJ}$, 
the denominator contains the Goldreich current for the axisymmetric case, while in 
the case of nonzero angles $\chi$, the Goldreich-Julian charge density in the vicinity of 
the magnetic poles
\begin{equation}
\rho_{\rm GJ} = - \frac{{\bf \Omega} {\bf B}}{2 \pi c} \approx - \frac{\Omega B}{2 \pi c} \, \cos\chi 
\label{GJ}
\end{equation}
itself depends on the angle $\chi$. On the other hand, it is natural to expect the longitudinal 
current to be bounded by the value $j_{\parallel} \approx \rho_{\rm GJ} c$. At any rate, both 
in the Ruderman-Sutherland model [14] with the particle escape from the surface of a neutron 
star being hindered and in the Arons model [46], in which the escape of particles is free, the 
longitudinal current is indeed determined by the relation $j_{\parallel} \approx \rho_{\rm GJ} c$. 

But then, in the case of an oblique rotator, the dimensionless current $i_0$ has the upper bound 
$i_0^{(\rm max)}(\chi) \sim \cos\chi$. As a result, the current losses must decrease as 
the angle $\chi$ increases, at least like  $\cos^2\chi$. In particular, if  $\chi = 90^{\circ}$ 
(when $\cos^2\chi$ is to be substituted by its characteristic value in the range of the polar cap,  
$<\cos^2\chi> \approx \Omega R/c$), we obtain
\begin{equation}
W_{\rm tot} = c_{\perp}\frac{B_0^2\Omega^{4}R^6}{c^3}
\left(\frac{\Omega R}{c}\right) i_{\rm A}.
\label{e17'}
\end{equation}
In the case of a local Goldreich current, $i_{\rm A} = 1$, while the coefficient 
$c_{\perp} \sim 1$ already depends not only on the profile of the asymmetric longitudinal current 
but also on the shape of the polar cap.

In discussions of this issue, the following reasoning is standardly used as an argument against 
the decrease in losses occurring as $\chi$ increases. In expression (3) for the decelerating 
moment, an increase in the angle $\chi$ is indeed accompanied by the surface current $J_{\rm s}$ 
decreasing as $\cos \chi$. But the characteristic distance between the axis and the polar cap 
increases as $\sin \chi$, and hence, even in the case of a local Goldreich current, the losses 
depend weakly on the inclination angle.

However, as has been demonstrated by a more precise analysis in Ref. [1], the above reasoning, 
which seems obvious at first glance, does not take the real structure of surface currents in 
the polar cap region into account. As shown in Fig. 3, the currents that close up should actually 
be arranged such that the current averaged over the polar cap surface vanish. Consequently, in 
determining the deceleration rate of a radio pulsar, it is necessary to consider higher-order 
effects, such as the effect of the curved surface of a neutron star.

But if the averaged surface current within the polar cap indeed vanishes, then, as shown in 
Fig. 3, a surface current comparable in value to the volume current flowing in the magnetosphere 
should flow along the separatrix separating the regions of closed and open field lines. In the 
case of an orthogonal rotator (and for a circular polar cap, when the result can be obtained 
analytically), the inverse current should amount to $3/4$ of the volume current flowing in the 
region of open field lines. A remarkable event was that numerical simulation [47] of the 
magnetosphere of an oblique rotator actually revealed inverse currents flowing along the 
separatrix. True, the inverse current only amounted to 20\% of the volume current. But such a 
discrepancy between the results of simulation and theoretical predictions can most likely be 
explained by the radius of the star being set in simulations only to half the radius of the 
light cylinder, when the magnetic field near the magnetic poles already differs strongly from 
the dipole field.

\begin{figure}
\begin{center}
\includegraphics[width=6cm]{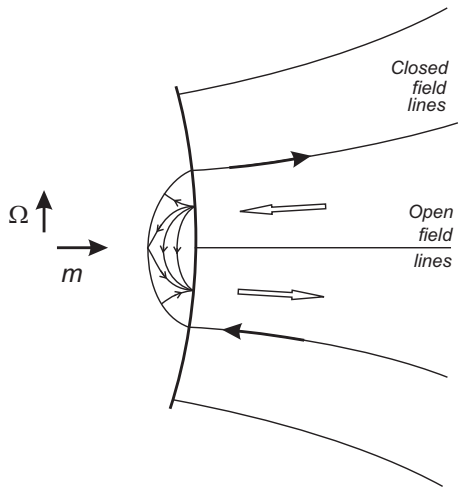}
\end{center}
\caption{\small{Structure of electric currents flowing in the vicinity of the magnetic poles 
of an orthogonal rotator. The currents flowing along separatices (bold arrows), separating 
the regions of closed and open field lines, close the longitudinal volume currents (contoured 
arrows) such that surface currents are fully concentrated within the polar cap.}
   }
\label{fig1_3}
\end{figure}

Finally, we note that no contradiction actually exists, either, between relation (4) and the 
expression
\begin{equation}
W_{\rm tot}  \approx \frac{1}{4} \, \frac{B_0^2\Omega^4R^6}{c^3}\left(1 + \sin^2\chi\right),
\label{Wspit}
\end{equation}
obtained by Spitkovsky for an oblique rotator; approximate formula (9) was obtained in Ref. [20] 
for a magnetosphere in which the longitudinal current was actually significantly larger than the 
local Goldreich current (see Ref. [48] for the details), which is consistent with the condition 
$i_0 > 1$ (correspondingly, $i_{\rm A} > 1$). A longitudinal current exceeding the local Goldreich 
current was necessary for constructing a smooth solution containing the magnetohydrodynamic (MHD) 
wind outflowing to infinity (see Section 2.4). 

It is interesting that one more possibility was recently revealed for directly testing the 
validity of expression (4) for current losses $W_{\rm tot}$ (and at the same time the validity 
of the assertion that the pulsar magnetosphere completely screens the magnetodipole radiation 
of a neutron star). The possibility of implementing such a test is related to the unusual 
properties of the pulsar B1931$+$24 [49]. Unlike the radiation of other radio pulsars, 
the radiation of B1931$+$24 is strongly variable. This pulsar is in an active state for 
$5$--$10$ days, then its radio emission is switched off in less than 10 s, and it is no longer 
observable for the next $25$--$30$ days. It is important that the absolute value of the rotation 
deceleration of B1931$+$24 is different in the 'on' and 'off' states:
\begin{eqnarray}
\dot \Omega_{\rm on} & = & - 1.02 \times 10^{-14} \, [{\rm s}^{-2}], \\
\dot \Omega_{\rm off} & = & - 0.68 \times 10^{-14} \, [{\rm s}^{-2}],
\end{eqnarray} 
with 
\begin{equation}
\frac{\dot \Omega_{\rm on}}{\dot \Omega_{\rm off}} \approx 1.5.
\end{equation}
Later, the pulsar J1832$+$0031 was also found to exhibit similar behavior 
($t_{\rm on} \sim 300$ days, $t_{\rm off} \sim 700$ days, and in this case, 
also, the ratio $\dot \Omega_{\rm on}/\dot \Omega_{\rm off} \approx 1.5$),
as did the pulsar J1841$-$0500 (in that case, 
$\dot \Omega_{\rm on}/\dot \Omega_{\rm off} \approx 2.5$ [50]).

It is natural to assume that the difference between $\dot\Omega_{\rm on}$ and 
${\dot\Omega_{\rm off}}$ for these pulsars occurs simply because deceleration 
in the switched-on state is related to current losses, while in the switched-off 
state (when the magnetosphere is not filled with plasma), it is due to the emission 
of a magnetodipole wave [51, 52]. Then, using relations (1) and (4), we obtain
\begin{equation}
\frac{\dot\Omega_{\rm on}}{\dot\Omega_{\rm off}} 
= \frac{3f_{\ast}^2}{2} \, {\rm cot}^2\chi,
\end{equation}
which yields a reasonable value for the inclination angle 
$\chi \approx 60$--$70^{\circ}$. On the other hand, if relation (9) is 
applied for the switched-on state, we arrive at
\begin{equation}
\frac{\dot\Omega_{\rm on}}{\dot\Omega_{\rm off}} = \frac{3}{2} \,
\frac{(1+ \sin^2\chi)}{\sin^2\chi}.
\end{equation}
Clearly, this quantity cannot be equal to 1.5 or 2.5 for any value of the inclination 
angle{\footnote{For this reason, relation (9) was somewhat corrected in Ref. [53].}}. 
If such an interpretation of the observations corresponds to reality, it follows that 
the longitudinal current flowing in the magnetosphere does not actually exceed the 
local Goldreich current.

\subsection {Evolution of the inclination angle}

Determining the evolution of the inclination angle $\chi$ could serve as one more test. For 
current losses (more exactly, for local Goldreich current, and for inclination angle
$\chi \neq 90^{\circ}$), the decelerating moment ${\bf K}$ is directed opposite to the magnetic 
moment of the neutron star ${\bf m}$, and therefore the Euler equation leads to the conservation 
of the projection of the rotation angular velocity ${\bf \Omega}$ onto the axis perpendicular 
to ${\bf K}$. Hence, the following quantity must be conserved during the evolution [3]:
\begin{equation}
\Omega\sin\chi = {\rm const}.
\label{c185}
\end{equation}
Consequently, in the case of current losses, the angle $\chi$ between the axis of rotation 
and the magnetic axis should increase (but not decrease, as in the case of magnetodipole 
radiation), and the characteristic time of its evolution should coincide with the 
characteristic dynamical age of the pulsar,  $\tau_{\rm D} = P/2\dot P$ [34]. 
Regretfully, no method has been found to determine the direction of evolution of the inclination 
angle $\chi$ for individual pulsars (see, however, Ref. [54]). On the other hand, the prediction 
indicating an increase in the angle $\chi$ is known not to contradict observations statistically 
[34, 55].

The last assertion requires clarification. Observations reveal average statistical 
inclination angles $\chi$ indisputably {\it decrease} as the period $P$ of pulsars 
increases and its derivative $\dot{P}$ decreases [56]. Therefore, the average 
statistical inclination angle  decreases as the dynamic age $\tau_{\rm D}$ increases. 
Correspondingly, pulsars with larger periods can be observed to exhibit relatively 
larger widths of the mean pulses $W_{\rm r} = W_{\rm r}^{(0)}/\sin\chi$ [57] (where 
$W_{\rm r}^{(0)}$ is the width of thedirectivity pattern). But this by no means 
implies that the inclination angle for each concrete pulsar decreases with time. 
Such a behavior of the average inclination angle $\chi$ can also be realized when 
the angles $\chi$ of each pulsar increase in accordance with (15).

\begin{figure}
\begin{center}
\includegraphics[width=\columnwidth]{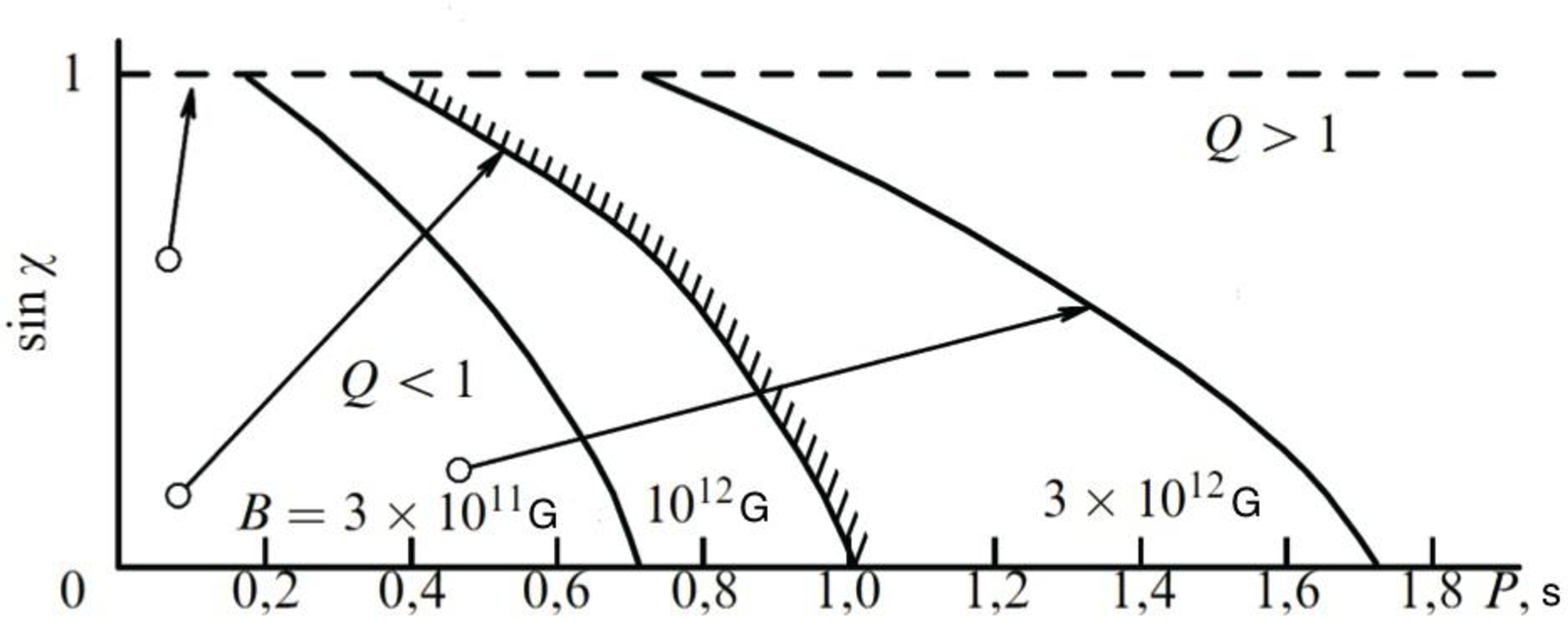}
\end{center}
\caption{\small{Pulsar extinction line in a $P$--$\sin\chi$ diagram for different inclination 
angles $\chi$ and different magnetic fields. Arrows show the evolution tracks of individual pulsars 
in the model of the current losses (15). The production of secondary particles is suppressed 
at angles  $\chi$ close to $90^{\circ}$. Therefore, neutron stars, which in the diagram are 
above and to the right of the extinction line, do not manifest themselves as radio pulsars, 
independently of the deceleration mechanism.}
   }
\label{fig1_4}
\end{figure}

Indeed, as can be seen from Fig. 4, for the given values of the pulsar period $P$ and the 
magnetic field $B_{0}$, the production of particles is suppressed precisely at angles $\chi$ 
close to $90^{\circ}$. This is because Goldreich-Julian charge density (7) decreases 
significantly at such angles, which in turn leads to a decrease in the electric potential drop 
near the surface of the neutron star. As a result, stable production of secondary 
particles becomes impossible there. Therefore, owing to such a dependence of the pulsar 
extinction line on $\chi$, the average inclination angle can also decrease 
as the dynamic age increases, for example, in the case of pulsars uniformly distributed over 
the $(P, \sin\chi)$ plane. A detailed analysis, already carried out in Ref. [34] on the basis 
of a kinetic equation describing the distribution of pulsars (see also later studies [58, 59]) 
confirmed that the observed distribution of pulsars over the inclination angle does not 
contradict hypothesis (15) of the increase in the angle $\chi$ for any individual 
pulsar.

In any case, it is quite clear, that the inclination angle $\chi$ is a key hidden parameter: 
without taking it into account, it is impossible to construct a consistent theory of the 
evolution of radio pulsars. Regretfully, with a few exceptions (see, e.g., Ref. [60]), modern 
theorists (so-called scenario producers) describing the evolution of neutron stars [61-63] do 
not take the influence of the inclination angle evolution on the observed pulsar distribution 
into account.

\subsection {Light surface}

Starting from the 1970s, the magnetosphere of a pulsar has been discussed almost exclusively 
in the force-free approximation [23, 64-66]. The reason is that the plasma filling the 
magnetosphere of a neutron star is less important (secondary) than the magnetic field; 
therefore (at least within the light cylinder), the particle energy density can be neglected.
 
In the force-free approximation, the structure of the magnetosphere is described by the so-called 
pulsar equation, i.e., an elliptic equation for the magnetic flux function. In Section 2.1, in 
discussing the solution of a similar equation for the zero longitudinal current, we noted that 
in the case of numerical simulation of an axisymmetric magnetosphere, it is necessary to introduce 
an additional condition for the external boundary of the integration region [24-29]. Such a 
condition is usually chosen in the form that the magnetic field lines be radial. In this approach, 
precisely such an additional condition fixes the longitudinal current flowing within the 
magnetosphere. Therefore, it is not surprising that the current is close to the critical 
current $j_{\rm GJ} = \rho_{\rm GJ} c$ (5) obtained analytically by F C Michel [64] for the 
quasispherical wind.

A very important property of this solution is that the energy in the wind is carried by the 
crossed fields  $E_{\theta}$ and $B_{\varphi}$, which form a radial flux of the electromagnetic
energy (the Poynting vector flux), and the electric field is smaller than the magnetic 
field as far as infinity. Otherwise, the freezing-in condition 
\mbox{${\bf E} + {\bf v} \times {\bf B}/c = 0$,} which always serves as the cornerstone 
in the approach considered, would be violated.

On the other hand, as is readily verified, implementation of the condition $E < B$ is possible 
only if the longitudinal electric current $I$ flowing in the magnetosphere 
is suffiently large. Indeed, in the case of a quasispherical wind outside the light cylinder, 
the electric field
\begin{equation}
E_{\theta} = \frac{\Omega r \sin\theta}{c} B_{\rm p}
\label{Etheta}
\end{equation}
and the toroidal magnetic field
\begin{equation}
B_{\varphi} = \frac{2I}{cr \sin\theta}  
\label{Bphi'}
\end{equation}
decrease with the distance  $r$ as $r^{-1}$ (while the poloidal field $B_{\rm p}$ 
decreases as $r^{-2}$). Therefore, for the light surface to recede to infinity, it 
is necessary that the toroidal magnetic field on the light cylinder be of the same 
order of magnitude as the poloidal field. Implementation of this condition is possible 
precisely when the total current $I$ outflowing beyond the light cylinder is not less than 
the Goldreich current $I_{\rm GJ} = \pi j_{\rm GJ} R_{0}^2$, where $R_{0}$ is the radius 
of the polar cap. We stress that in all numerical simulations, no restrictions were 
imposed on the longitudinal current outflowing from the neutron star surface. Therefore, 
it is not surprising that the longitudinal current obtained as a solution of the problem 
turned out to be of the order of $I_{\rm GJ}$. 

We recall that in the complete MHD version, where taking the finiteness of particle masses 
into account results in the appearance of an additional critical (fast magnetosonic) surface, 
the longitudinal current is no longer a free parameter [67]. The value of the longitudinal 
current is close to $I_{\rm GJ}$. Therefore, most reseachers currently consider the existence 
of a strongly magnetized wind for which the condition $|{\bf E}| < |{\bf B}|$ is satisfied to 
be practically proven [20, 21]. We stress that the issue here concerns scales comparable to the 
radius of the light cylinder ($r \sim 1$--$100 \, R_{\rm L}$); at larger scales, as follows, 
for instance, from an analysis of the interaction of the pulsar wind with supernova remnants [68], 
the main energy must already be carried by particles. As is known, within the theory of strongly 
magnetized wind, such an acceleration required for a quasispherical outflow cannot be obtained 
[67, 69-71].

Generally speaking, the rigorous analytic conclusion that the longitudinal current is close to 
the critical one only concerned stationary axisymmetric flows. But numerical calculations 
recently performed for nonstationary force-free configurations [53] have shown unambiguously 
that the system undergoes quite rapid evolution precisely toward a state with a current close 
to the critical current. And such a state corresponds to the minimum-energy configuration (for 
example, minimum energy is exhibited by configurations in which the singular point separating 
the regions of closed and open field lines is on the light cylinder, but not inside the 
magnetosphere [29, 72]{\footnote{Solutions in which the singular point is inside the light 
cylinder are, most likely, affected by the limited time available for numerical calculations 
(A Tchekhovskoy, private communication).}}). Thus, the existence of quite a strong longitudinal 
current has been confirmed once again; in any case, no restrictions were imposed on the 
longitudinal current either.

The following problem arises here, however. As noted, all the theories of stationary particle 
production in the magnetosphere of a neutron star [3, 14, 46, 73, 74] unambiguously testify in 
favor of the longitudinal current density not possibly being larger than the local Goldreich 
current, which, as can be seen from its definition (7), depends on the inclination angle $\chi$
\begin{equation}
j_{\rm GJ} = \frac{ \Omega B}{2\pi}\cos\chi.
\label{GJforj}
\end{equation}
For example, in the case of an orthogonal rotator, the local Goldreich-Julian charge density in 
the vicinity of the magnetic poles must be $(\Omega R/c)^{1/2}$ times smaller than in the case of 
an axisymmetric magnetosphere. Hence, the longitudinal current flowing along open field 
lines should also be smaller in the same proportion (for ordinary pulsars with a period $P \sim 1$ 
s, this current is nearly 100 times smaller for an orthogonal rotator). Therefore, in constructing 
the solution for an oblique dipole [20], it was necessary to assume the longitudinal current in 
the region of the magnetic poles to be significantly higher than the local Goldreich current 
(A Spitkovsky, private communication).

Therefore, the value of the longitudinal current flowing in a neutron star's magnetosphere 
turns out to be the key issue, without resolving which it is impossible to move toward the 
understanding of the structure of the magnetosphere of radio pulsars. The problem lies in 
whether the region of plasma generation at the surface of a neutron star can provide the 
longitudinal current sufficiently large for the existence of an MHD wind from an oblique 
rotator. If the necessary current can be created (see, e.g., Ref. [75], where one-dimensional,
nonstationary regime was considered), nothing can prevent the production of the MHD wind in 
which the main part of the energy is carried by the electromagnetic field: such is the opinion 
of most reseachers. But if the generation of a sufficiently high longitudinal current turns out 
to be impossible for some reason, then, in the vicinity of the light cylinder, a 'light surface' 
inevitably arises in which the electric field becomes equal to the magnetic field. Precisely such 
a structure was predicted by us in Ref. [1].

\begin{figure}
\begin{center}
\includegraphics[width=8cm]{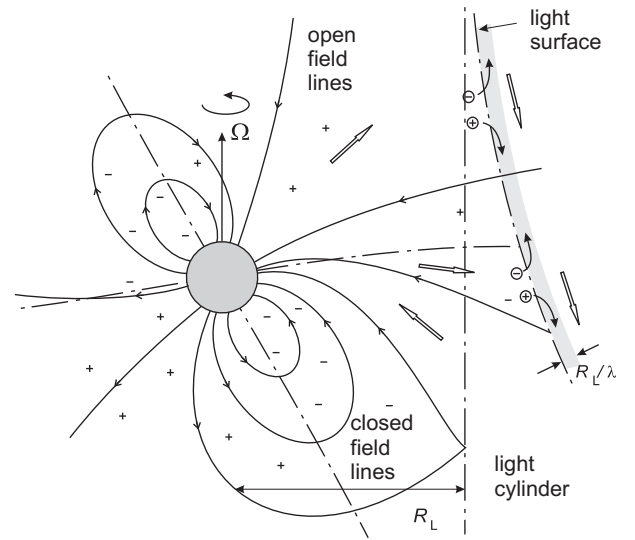}
\end{center}
\caption{\small{Structure of a magnetosphere with a small enough longitudinal current, having a 
natural boundary --- the 'light surface' $|{\bf E}|=|{\bf B}|$ at which the freezing-in 
condition cannot be satisfied. Therefore, electrons and positrons are accelerated in opposite 
directions along the electric field. Precisely this current closes the longitudinal currents 
flowing in the magnetosphere.}
   }
\label{fig1_5}
\end{figure}

The appearance of a light surface in the magnetosphere of a radio pulsar radically alters the 
properties of the pulsar wind because, close to the light surface, closure of the current and 
effective particle acceleration inevitably occur. Solving the equations of two-fluid hydrodynamics 
(precisely describing the difference in motion between electrons and positrons) in the case of 
the simplest cylindrical geometry reveals [1] that a significant part of the energy carried 
within the light surface by the electromagnetic field in the thin transition layer close to 
the light surface,
\begin{equation}
\Delta r \sim \lambda^{-1}R_{\rm L}
\label{drl}
\end{equation}
is transferred to plasma particles (\mbox{$\lambda = n_{\rm e}/n_{\rm GJ} \sim 10^3$--$10^5$} 
is the production multiplicity of particles close to the surface of a neutron star). In the 
same layer, as shown in Fig. 5, practically total closure of the longitudinal current circulating 
in the magnetosphere occurs. As a result, a natural explanation is also found for the high 
efficiency of particle acceleration. Subsequently, a similar result was also obtained for a 
more realistic geometry, when the poloidal magnetic field near the surface of the light cylinder 
is close to a monopole field [76]. In particular, it has been confirmed that the particle energy 
immediately beyond the light surface is by the order of magnitude given by
\begin{eqnarray}
{\cal E}_{\rm e} & \sim & eB_0 R\frac{1}{\lambda}\left(\frac{\Omega R}{c}\right)^2 
\label{Emax}
\\
& \sim & 10^4 \, {\rm MeV}
\left(\frac{\lambda}{10^3}\right)^{-1}
\left(\frac{B_0}{10^{12}{\rm G}}\right)
\left(\frac{P}{1{\rm s}}\right)^{-2},
\nonumber
\end{eqnarray}
but does not exceed the value $10^6$, at which the effects of radiation friction become essential.

Quantity (20) practically corresponds to the total energy transfer from the Poynting vector 
flux to the flux of accelerated particles. In (20), ${\cal E}_{\rm e}$ is $\lambda$ times 
smaller than the energy ${\cal E}_{\rm max} = e \Delta V$ corresponding to the maximum potentials
difference $\Delta V \sim (\Omega L/c) B L$ of different magnetic field lines in the 
magnetized wind. Here, $L$ is the characteristic size of the central engine; in radio pulsars, 
$L$ is equal to the size of the polar cap $R_{0} = (\Omega R/c)^{1/2}R$. As a result, we can 
express the Lorentz factor $\gamma_{\rm max} = {\cal E}_{\rm e}/m_{\rm e} c^2$ as 
\begin{equation}
\gamma_{\rm max} = \sigma,
\label{giss}
\end{equation}
where $\sigma$ is the so-called magnetization parameter{\footnote{Recently, the notation 
$\mu$ for the magnetization parameter has become popular, with $\sigma$ standing for the 
ratio of the electromagnetic energy flux to the energy flux carried by particles.}} introduced 
by Michel [77] in 1969. As was shown in Ref. [48], the magnetization parameter can be represented 
in the very simple form
\begin{equation}
\sigma 
\sim \frac{1}{\lambda}\left(\frac{W_{\rm tot}}{W_{\rm A}}\right)^{1/2},
\label{newsigma}
\end{equation}
where $W_{\rm tot}$ is the total energy losses and 
\mbox{$W_{\rm A} = m_{\rm e}^{2}c^{5}/e^{2} \approx 10^{17}$ 
erg s$^{-1}$} is a universal constant; it corresponds to the minimum energy losses
of the central engine which can accelerate particles to relativistic energy. Because 
the particle production multiplicity $\lambda \sim 10^{3}$--$10^{5}$ for radio pulsars 
is known [73, 74, 78], the value of $\sigma$ can also be found.For most pulsars, the 
value of $\sigma$  lies in the range $10^{3}$--$10^{4}$, and it can reach $10^{6}$ only 
in the youngest sources (the Crab and Vela pulsars). We note that the parameter $\sigma$ 
is very convenient for determining the key parameters of strongly magnetized winds. For 
example, the radius of the fast magnetsonic surface is expressed simply as
\begin{equation}
r_{\rm f} \sim \sigma^{1/3} R_{\rm L}. 
\label{rfms}
\end{equation}

Therefore, our theory predicts effective particle acceleration in the region of the light 
cylinder up to energies corresponding to the Lorentz factor $10^{4}$--$10^{6}$. 
Clearly, such an acceleration is only possible within the fast magnetosonic surface, 
$r < r_{\rm f}$; as was noted, if the magnetized wind is free to reach the fast magnetosonic 
surface, then the longitudinal current is comparable in value to the Goldreich current.

Clearly, the effective acceleration of particles should result in the generation of hard 
radiation, which, in principle, could be detected. In [3], the synchrotron losses of 
accelerated particles are estimated, and both the total energy release and the energy 
range of radiation are shown to depend very strongly on the period $P$ of the pulsar. The 
energies of emitted photons can reach several tens of MeV only in the case of the youngest 
pulsars (Crab, Vela), while the radiation of most radio pulsars due to the synchrotron 
mechanism must lie in the optical range. The energy released in all ranges has turned out 
to be quite low, which has allowed observing these sources with the aid of existing detectors.

On the other hand, as is well known, another important channel in which energy is released 
and which is capable of resulting in the production of $\gamma$-quanta of even higher energies 
is the inverse Compton scattering of thermal X-ray photons emitted from the surface of a 
neutron star. This process has recently been regarded as the main process for the generation 
of photons of energies in the TeV range for the widest class of objects, such as active galactic 
nuclei [79,80], galactic sources in the TeV range [81], and, naturally, radio pulsars [82]. In 
those cases where the 'central engine' is indeed a rapidly rotating neutron star, the Lorentz 
factor of electrons (or positrons) necessary for shifting the observed photons and soft
$\gamma$-quanta toward the TeV energies corresponds precisely to values $\gamma = 10^{4}$--$10^{5}$. 
Thus, for the pulsar B1259$-$63 (which is in a double system containing a Be-star), 
observations agree best with the value $\gamma \sim 10^{4}$ [83]. As can be readily estimated 
from relations (21) and (22), precisely this value is the characteristic value of the 
magnetization parameter $\sigma$ for B1259$-$63 . 

We especially draw attention to the work of the group of F Aharonian published in {\it Nature} 
[82], the title of which is precisely the following: "Abrupt acceleration of a 'cold' 
ultrarelativistic wind from the Crab pulsar." It is shown in Ref. [82], for example, that 
the observed intensity of TeV photons can be explained, if a rapid acceleration of particles 
occurs at distances of the order of  $30 \, R_{\rm L}$ as a result of which the particles 
acquire energies corresponding to Lorentz factors $\gamma \approx 10^{6}$. As was noted above, 
the value $\gamma \sim 10^{6}$ corresponds precisely to estimate (22) for the magnetization 
parameter $\sigma$ of the Crab pulsar. Moreover, the scale of $30 \, R_{\rm L}$ is certainly 
smaller than the radius of the fast magnetosonic surface $r_{\rm f} \approx 100 \,R_{\rm L}$
(see Eqn 23). 

A detailed comparison of theoretical predictions and observational data is beyond 
the scope of this article, however{\footnote{In our opinion, the distance from the 
acceleration region to the light cylinder amounting to $30R_{\rm L}$ for the Crab 
pulsar may be overestimated.}}. Nevertheless, it must be noted that after it became 
clear that the existence of a large potential difference $\Delta V$ between the magnetic 
field lines in the magnetized wind does not lead directly to any effective particle 
acceleration up to ultra-high energies ${\cal E}_{\rm max} \sim \lambda {\cal E}_{\rm e}$ 
(see, e.g., Ref. [84]), the possibility of direct electrostatic acceleration of particles 
is not being taken into consideration (see, e.g., Ref. [85]). However, as was shown above, 
this process could well be realized in the case where, for some reason or other, the 
longitudinal current flowing in the magnetosphere of a compact object is quite small.

\section{Theory of radio emission}

\subsection{Formulation of the problem}

As is well known, one of the methods for determining the instability increment of waves in 
a plasma consists in analyzing the dispersion equation, for which it is necessary to determine 
the dielectric permittivity tensor of the medium. The procedure for calculating the dielectric 
permittivity tensor of an inhomogeneous anisotropic plasma in the approximation of geometric 
optics on the basis of the standard approach, expounded, for example, in Ref. [86], is described 
in detail in [2]. In the same work, a study is presented of the collective interaction in which 
electromagnetic waves associated with curvature radiation are simultaneously amplified by the 
Cherenkov mechanism. We emphasize that this effect is absent in the vacuum. Most likely because 
the calculation procedure is quite complicated, erroneous assertions have been made in a number 
of publications [87-89]. The objections put forward in Ref. [89] were later withdrawn [90] by 
the author. As regards Refs [87, 88], which are still cited in papers on the relevant topic, 
they contain numerous arithmetic errors, which have all been revealed and described in detail 
in Ref. [42].

In addition, another method for dealing with the problem of collective curvature radiation was 
proposed in [36-38, 91]. In these studies, a model problem, which could be solved 'exactly', 
was considered in a cylindrical geometry. The magnetic field in this problem is assumed to be 
precisely circular, while the relativistic plasma moves along the circular magnetic field lines 
owing to the centrifugal drift, \mbox{$u = c\rho_c/\rho_0 \ll c$}, directed parallel to the 
cylindrical axis. Here, $\rho_c=c/\omega_{\rm c}$ is the cyclotron radius and $\rho_0$ is the 
curvature radius. But as we now show, this approach cannot be used in analyzing the collective 
curvature radiation either (in more detail see [96]).

As in [36-38], we consider the electromagnetic fields in the wave to be of the form
\begin{equation}
\left[{\bf E},{\bf B}\right] = \left[{\bf E}(\rho),{\bf B}(\rho)\right]\times
\exp\left\{-i \omega t+is\varphi+ik_z z\right\},
\end{equation}
where, $\omega$ is the wave frequency,  $s$ is an integer defining the azimuthal component
\mbox{$k_\varphi = s/\rho$} of the wave vector, and  $k_z$ is the component of the wave 
vector parallel to the cylinder axis. In the approach considered, the amplitudes are 
assumed to depend only on the coordinate $\rho$. Moreover, not the vectors ${\bf E}$ and 
${\bf B}$, but their cylindrical components $(E,B)_{\rho}, (E,B)_{\varphi}$ and $(E,B)_{z}$ 
depend only on the cylindrical radius $\rho$. This means that the polarization in the wave 
follows the magnetic field, turning from one point to another, which is possible only in the 
case of a well-defined boundary condition, for instance, for a system inside a metal cylinder. 
Under these conditions, we arrive at a one-dimensional problem, which can indeed be readily 
solved. However, as is not difficult to show, the wave considered within such an approach has 
nothing to do with curvature radiation.

To show this, we consider a particle moving along a circular trajectory of radius $\rho_0$ with a 
constant velocity  $v$. Such motion corresponds to an infinite magnetic field. Then the emitted 
energy is equal to the work performed by the field of the wave on the electric current of the 
particle. The electric current is given by
\begin{equation}
{\bf j}= e v \delta(\varphi-\Omega t)\, \delta(z)
\frac{\delta(\rho-\rho_0)}{\rho}\, {\bf e}_\varphi,
\end{equation}
where $\Omega=v/\rho_0$. For the polarization chosen,
\begin{equation}
\int {\bf jE} \, {\rm d}{\bf r}=evE_\varphi(\rho_0)\exp\left\{-i\omega t+is\Omega t\right\}.
\end{equation} 
It hence follows that radiation is possible only when $\omega=s\Omega=k_\varphi v$. 
But this is the condition for Cherenkov, but not curvature, radiation. A wave with such 
polarization cannot be generated by the curvature mechanism. The difference between curvature 
and Cherenkov waves is that the interaction time of bremsstrahlung and the irradiating particle 
is finite. A freely propagating wave with a nearly constant polarization deviates from the 
direction of motion of the particle. As a result, a nonzero projection of the electric field 
of the wave arises in the direction of the electric current of the particle, i.e., an exchange 
of energy between the wave and the particle becomes possible. The process lasts a finite time 
$\tau$, which can be found from the condition $\tau(\omega-{\bf kv})\simeq 1$. For relativistic 
particles ($v\simeq c$) we have $\tau=(\rho_0^2/\omega c^2)^{1/3}\simeq \rho_0/c\gamma$.

\subsection{Polarization of the curvature wave}

It is well known that the radiation field of the electric current ${\bf j}$ and of the charge 
density $\rho_{\rm e}$ of a moving particle of charge $e$ is described by retarded potentials 
[43]
\begin{eqnarray}
{\bf A} & = & \frac{1}{c}\int\frac{{\bf j}(t')}{R}{\rm d}{\bf r},
\label{A}\\
\Phi & = & \int\frac{\rho_{\rm e}(t')}{R}{\rm d}{\bf r},
\label{Q}
\end{eqnarray}
where $t'=t-R/c$ is the so-called retarded time and $R$ is the distance from the charge to the 
observer who is at the point with coordinates $(\rho, \varphi, z)$ at the moment $t'$,
\begin{eqnarray}
R & = & \left[\rho^2+z^2+\rho_0^2-2\rho\rho_0\cos(\varphi'-\varphi)\right]^{1/2}, \\
\varphi' & = & \Omega t'.
\end{eqnarray}
We now introduce the Fourier transform of potentials (27) and (28) 
with respect to time:
\begin{eqnarray}
{\bf A}_\omega & = & \frac{1}{2\pi}\int{\bf A}(t)\exp\{i\omega t\}{\rm d}t, \\
\Phi_\omega & = & \frac{1}{2\pi}\int\Phi(t)\exp\{i\omega t\}{\rm d}t.
\end{eqnarray}
It is convenient in what follows to pass from integration over $t$ to integration over the retarded 
time $t'$, and then over the angle  $(\varphi'-\varphi)$. As a result, in Cartesian coordiates 
($x,y,z$) we obtain the vector potential ${\bf A}$ and the scalar potential  $\Phi$ 
in the form
\begin{equation}
\left[{\bf A}_{\omega};\,\Phi_{\omega}\right]=\frac{e\rho_0}{2\pi c}\exp\{i\omega\varphi/\Omega\}
\left[-K_s;\,K_c;\, 0;\,\frac{c}{v}K_0\right],
\label{eqn1}
\end{equation}
where $K_0$, $K_{\rm s}$, and $K_{\rm c}$ are functions of only the coordinates $\rho$ and $z$:
\begin{eqnarray}
K_0 & =& \int\frac{\exp\{i\omega(R/c+\Omega^{-1})\alpha\}}
{R+v\rho\sin\alpha/c}{\rm d}\alpha, \nonumber \\
K_{\rm s} & = & \int\frac{\exp\{i\omega(R/c+\Omega^{-1})\alpha\}\sin\alpha}
{R+v\rho\sin\alpha/c}{\rm d}\alpha, \\
K_{\rm c} & =& \int\frac{\exp\{i\omega(R/c+\Omega^{-1})\alpha\}\cos\alpha}
{R+v\rho\sin\alpha/c}{\rm d}\alpha, \nonumber \\
R & = & (\rho^2+z^2+\rho_0^2-2\rho\rho_0\cos\alpha)^{1/2}. \nonumber
\end{eqnarray}
We stress that expression (33) is valid not only in the wave zone but also at any point ${\bf r}$.
The dependence on the angle  $\varphi$ is given by the factor $\exp\{i\omega\varphi/\Omega\}$. 
Hence, owing to the periodicity in the angle $\varphi$ we simply obtain  $\omega=s\Omega$.

An important fact follows from relation (33): the field of the irradiated wave is a superposition 
of three harmonics:  $s$, $s-1$ and $s+1$. For example, we present the following expression for the
azimuthal component of the electric field $E_{\varphi\omega}$:
\begin{eqnarray}
E_{\varphi\omega}  =  \frac{i\omega}{v}\left(-\frac{\rho_0}{\rho}\Phi_\omega
+\frac{v}{c}A_{\varphi\omega}\right) = 
\label{eqn6} 
\end{eqnarray}
$$-i\frac{e\rho_0\omega}{2\pi v^2}\exp\{is\varphi\}\left[
\frac{\rho_0}{\rho}K_0-\frac{v^2}{c^2}\left(K_{\rm s}\sin\varphi + K_{\rm c}\cos\varphi\right)
\right].$$
The first term in the right-hand part of (35), which is proportional to the scalar potential  
$\Phi$, is not significant in the wave zone $\rho \gg \rho_0$, but is significant in the 
vicinity of the particle trajectory, $\rho = \rho_{0}$. Owing to the presence of this term, 
a particle which is in resonance with one of the three harmonics, for instance, with $s$th 
($\omega = s\Omega$), is knocked out of resonance by the adjacent harmonics $s\pm 1$. Here, 
the component $E_{\varphi\omega}$ changes sign in a time $\tau$. The synchronicity condition 
$1-\cos\Omega\tau\simeq 1-v^2/c^2=\gamma^{-2}$ determines the time as $\tau$
\begin{equation}
\tau\simeq 1/\Omega\gamma=\rho_0/c\gamma,
\end{equation}
which coincides with the formation time of curvature radiation.

Hence, the emitted curvature wave comprises three harmonics, $s$ and $s \pm 1$, with a fixed 
relation between the amplitudes. Precisely this circumstance provides the curvature mechanism 
of radiation. The neighboring harmonics, $s\pm 1$, arise owing to modulation of the field of the 
emitted wave by the electric current of the particle whith the harmonic $s=1$. It 
can now be understood why simply dealing with collective curvature radiation in a cylindrical 
geometry with a single cylindrical harmonic does not reveal any significant amplification of 
waves [36-38, 91]. The chosen polarization excludes the curvature radiation.

\subsection{Propagation of the triplet of harmonics}

It was shown in Section 3.2 that the curvature radiation of a single charged particle in the 
vacuum cannot be described by a single cylindrical harmonic $\exp\{is\varphi\}$. In the problem 
of the collective curvature radiation of waves, modulation of the electric current of particles 
occurs at the same time as their excitation; therefore, the resonance azimuthal harmonic 
$s=\omega\rho/v_\varphi$ mixes with the harmonics of the modulation of the particle electric 
current, giving rise to harmonics with all possible values of $s$. Below, we show that all 
azimuthal harmonics contribute to the response of the medium to the electromagnetic field. 
But here, we show that propagation of the triplet $(s, s\pm 1)$ of cylindrical harmonics 
differs significantly from the propagation of a single harmonic, which is usually discussed 
in the literature.

We consider the simple cylindrical one-dimensional problem of the emission from a flux of cold 
relativistic plasma particles with charge $e$ and mass $m_{\rm e}$, which move in the $xy$ plane 
along an infinite azimuthal magnetic field $B_0 = B_{\varphi}$. In this case, the particles can 
only move in the $\varphi$-direction with a velocity $v_{\varphi}$ at an arbitrary cylindrical 
radius $\rho$. We assume the nonperturbed particle number density $n_{\rm e}^{(0)}$ and velocity 
$v_\varphi^{(0)}$ to be constant, i.e., independent of the radius $\rho$. The electric current 
${\bf j}$ then has only a component along $\varphi$, while the magnetic field of the wave has 
only the component ($B_\rho = B_\varphi=0$). The electric field in the wave has two nonzero 
components, $E_\rho$ and $E_\varphi$ ($E_z=0$).

The dependence of the wave field on the coordinates is given by
\begin{equation}
[E_{\rho}; \, E_{\varphi}; \, B_{z}] = [E_{\rho}(\rho); \, E_{\varphi}(\rho); 
\, B_{z}(\rho)]\exp\{-i\omega t +is\varphi\}.
\end{equation}
From the Maxwell equations, we obtain
\begin{eqnarray}
\frac{{\rm d} E^{(\sigma)}_{\varphi}}{{\rm d} \rho} = \frac{i \sigma}{\rho} E^{(\sigma)}_{\rho} -
i \frac{\rho}{\sigma}\frac{\omega^2}{c^2} E^{(\sigma)}_{\rho} - \frac{E^{(\sigma)}_{\varphi}}{\rho},\\
\frac{{\rm d} E^{(\sigma)}_{\rho}}{{\rm d} \rho} = - i \frac{\sigma}{\rho} E^{(\sigma)}_{\varphi} +
\frac{4 \pi}{\omega}\frac{\sigma}{\rho} j^{(\sigma)}_{\varphi} - \frac{E^{(\sigma)}_{\rho}}{\rho},
\end{eqnarray}
where the index $\sigma$ corresponds to one of the three harmonics, $s$ or $s\pm 1$.
For simplicity, we introduce the dimensionless variable $r = \rho\omega/c$, and the quantity
\begin{equation}
\Lambda = \frac{\omega^2_{\rm p}}{\omega^2\gamma^3}, 
\end{equation}
where $\omega_{\rm p} = (4\pi n_{\rm e}e^2/m_{e})^{1/2}$ is the plasma frequency,
\mbox{$\gamma=(1-v_\varphi^2/c^2)^{-1/2}$} is the Lorentz factor of plasma particles,
and
\begin{equation}
G_{\sigma} = \frac{4 \pi j^{(\sigma)}_{\varphi}}{\Lambda \omega}
\end{equation}
is the dimensionless current. In the new variables, Eqns (38) and (39) become
\begin{eqnarray}
\frac{{\rm d} E^{(\sigma)}_{\varphi}}{{\rm d} r} & = & \frac{i \sigma}{r} E^{(\sigma)}_{\rho} - i \frac{r}{\sigma}
E^{(\sigma)}_{\rho} - \frac{E^{(\sigma)}_{\varphi}}{r},
\label{eq78}
\\
\frac{{\rm d} E^{(\sigma)}_{\rho}}{{\rm d} r} & = & - i \frac{\sigma}{r} E^{(\sigma)}_{\varphi} +
\Lambda \frac{\sigma}{r} G_{\sigma} - \frac{E^{(\sigma)}_{\rho}}{r}.
\end{eqnarray}

As was noted, we here consider the interaction of three waves, $s$ and $s\pm 1$. It is very 
important that the propagation of these waves is not independent: coupling between the waves is 
realized by means of the electrostatic field $[E_\rho(\rho); \, E_\varphi(\rho)] \exp\{i\varphi\}$ 
with the lowest azimuthal harmonic $s=1$. The electrostatic field turns out to be the result of 
nonlinear coupling of the high-frequency harmonics $s$ and $s\pm 1$. The propagation equations of 
the mode $s = 1$ in the same notation have the form
\begin{eqnarray}
\frac{{\rm d} E_{\varphi}}{{\rm d} r} & = & \frac{i}{r} E_{\rho} - \frac{E_{\varphi}}{r},
\label{ref1} \\
\frac{{\rm d} E_{\rho}}{{\rm d} r} & = & - i \frac{1}{r} E_{\varphi} 
+ \Lambda Z - \frac{E_{\rho}}{r},
\label{eq90}
\end{eqnarray}
where  $Z = 4 \pi n_{\rm e} e c/(\Lambda \omega)$. We stress that the second term in the right-hand part 
of Eqn (42) is absent in Eqn (44) because the field proportional to $\exp(i\phi)$ is static 
\mbox{($\omega=0$).}

\begin{figure*}
\begin{center}
\includegraphics[width=14cm]{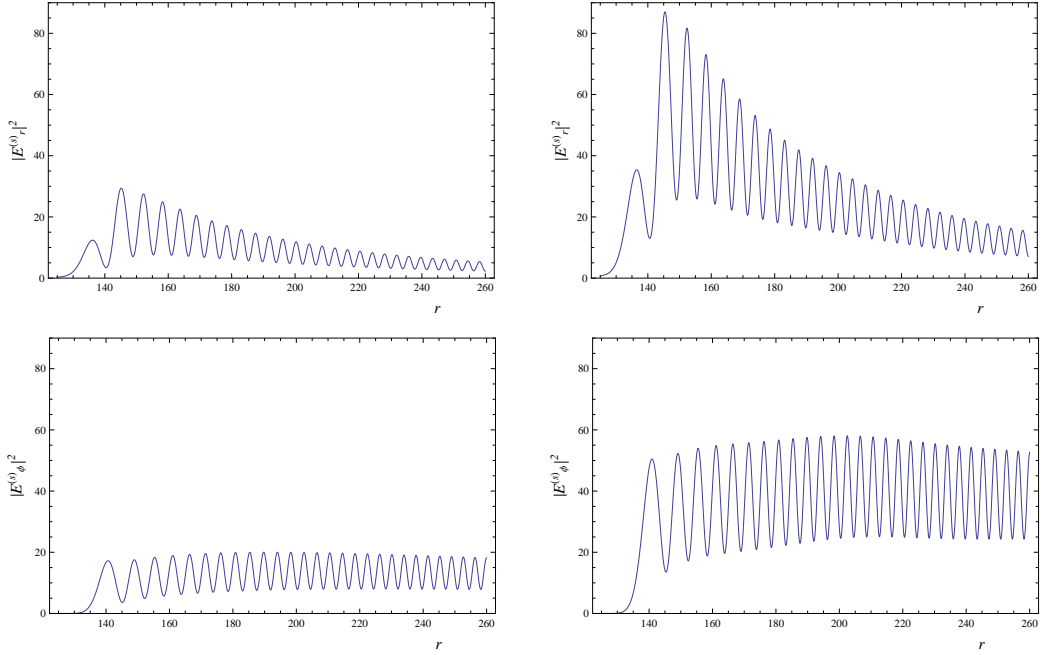}
\end{center}
\caption{\small{Wave amplification when the coupling the other harmonics is neglected (left), 
and when it is taken into account (right); $\Lambda = 10^{-2}$, $\nu = 1$ GHz, $\gamma =5$, 
and $s = 125$.}
   }
\label{fig2_1}
\end{figure*}

To determine the response of the medium to the electromagnetic fields, we can use the continuity 
equation and the Euler equation:
\begin{eqnarray}
&&\frac{\partial n_{\rm e}}{\partial t}+{\rm \nabla}(n_{\rm e} {{\bf v}})=0,\\
&&\left(\frac{\partial}{\partial t}+{\bf v}\nabla\right){\bf p}
=e\left({\bf E}+\left[\frac{{\bf v}}{c},{\bf B}\right]\right).
\end{eqnarray}
It is easy to verify that only the $\varphi$-component of the Euler equation is significant, 
while the radial component provides the equilibrium configuration across the infinite magnetic 
field.

We represent the velocity of plasma particles and the flux number density in terms of the expansion in 
powers of the wave field amplitudes:
\begin{eqnarray}
v_{\varphi} = v^{(0)}_{\varphi} +  \delta v^{(1)}_{\varphi} + \delta v^{(2)}_{\varphi} + ...,\\
n_{\rm e} = n_{\rm e}^{(0)} +  \delta n_{\rm e}^{(1)} + \delta n_{\rm e}^{(2)} + ....
\end{eqnarray}
The linear response can be readily found as
\begin{eqnarray}
n_{\rm e}^{(1)} & = & n_{\rm e}^{(0)} \frac{k v^{(1)}_{\varphi}}{\omega-kv_\varphi^{(0)}},
\label{denom1} \\
v^{(1)}_{\varphi} & = & i \frac{e E_{\varphi}}{m_{\rm e} \gamma^3 (\omega-kv_\varphi^{(0)})},
\label{denom2}
\end{eqnarray}
where $k=s/\rho$. To find the nonlinear response of the medium, it is necessary to take the 
nonlinear relation between  $\delta v_{\varphi}$ and $\delta p_{\varphi}$ into account:
\begin{equation}
\delta p_{\varphi} = m_{\rm e}\gamma^3 \delta v_{\varphi} -
\frac{3}{2}m_{\rm e} v^{(0)}_{\varphi}\gamma^5 \frac{(\delta v_{\varphi})^2}{c^2}.
\end{equation}
Straightforward calculation gives
\begin{eqnarray}
G_s & = & \frac{1}{1- s v^{(0)}_{\varphi}/r} \left[i\frac{E^{s}_{\varphi}}
{1- s v^{(0)}_{\varphi}/r}  + \alpha\frac{r}{v^{(0)}_{\varphi}} \right. 
\nonumber \\
& \times & 
\left(A_{s,s-1}\frac{E^{s-1}_{\varphi}E^{1}_{\varphi}}{1- (s -1)
v^{(0)}_{\varphi}/r} \right. 
\nonumber \\
& - & \left. \left. A_{s,s+1}
\frac{E^{s+1}_{\varphi}E^{1*}_{\varphi}}{1- (s +1) v^{(0)}_{\varphi}/r}\right)\right],
\label{Jeq1}\\
G_{s-1} & = & \frac{1}{1- (s -1) v^{(0)}_{\varphi}/r} 
\left[i\frac{E^{s-1}_{\varphi}}{1- (s -1) v^{(0)}_{\varphi}/r} \right.
\nonumber \\
& - & \left.\alpha\frac{r}{v^{(0)}_{\varphi}}A_{s,s-1}
\frac{E^{s}_{\varphi}E^{1*}_{\varphi}}{1- s v^{(0)}_{\varphi}/r}\right], \\
G_{s+1} & = & \frac{1}{1- (s +1) v^{(0)}_{\varphi}/r} 
\left[i\frac{E^{s}_{\varphi}}{1- (s +1) v^{(0)}_{\varphi}/r} \right.
\nonumber \\
& + & \left.\alpha\frac{r}{v^{(0)}_{\varphi}}A_{s,s+1}\frac{E^{s}_{\varphi}E^{1}_{\varphi}}{1- 
s v^{(0)}_{\varphi}/r}\right],
\end{eqnarray}
\begin{eqnarray}
Z & = & \frac{1}{\left(v^{(0)}_{\varphi}\right)^2} \left[i\frac{E^{1}_{\phi}}{1/r} + \right. 
\nonumber \\
& + & \alpha \left(\frac{E^{s+1}_{\varphi}E^{s*}_{\varphi}}{(1- (s +1) 
v^{(0)}_{\varphi}/r)(1- s v^{(0)}_{\varphi}/r)}\right. 
\nonumber \\
& + & \left.\left. \frac{E^{s}_{\varphi}E^{(s-1)*}_{\varphi}}{(1- s v^{(0)}_{\varphi}/r)(1- (s -1) 
v^{(0)}_{\varphi}/r)}\right)\right],
\label{Jeq2} \\
A_{a,b} & = & \frac{1}{1- a v^{(0)}_{\varphi}/r} + \frac{1}{1- b v^{(0)}_{\varphi}/r} -3 \gamma^2,
\nonumber
\end{eqnarray}
where $\alpha = e/(m_{e}c\gamma^3\omega)$ and the velocity of plasma particles is expressed 
in terms of the speed of light $c$. Similar quantities can be found in Ref. [3] for plane waves. 
The equations were analyzed numerically with the initial condition 
$E^{(\sigma)}_{\varphi} = -J^{'}_{\sigma}(r), E^{(\sigma)}_{r} = i {\sigma} J_{\sigma}(r)/r$, 
corresponding to the normal mode in the vacuum for a cylindrical geometry; here, 
$J_\sigma(r)$ is the Bessel function of order $\sigma$. The singularity in Eqns (53)--(56) was 
smoothed out, as usual, by adding a small term  $+i \varepsilon$ to the resonance denominators 
in (50) and (51).

Equations (42)--(45) were numerically solved for  $\sigma = s$ and \mbox{$\sigma = s \pm 1$} 
at two different values of $G_{\sigma}$ and $Z$. In the first case, we neglected the nonlinear 
terms in (53)--(56), while the second case corresponded to the fully nonlinear problem. The 
results of calculations for both cases are presented in Fig. 6. To illustrate the influence 
of the nonlinear current better, we have chosen the amplitude of the modes $s - 1$ and $s + 1$ 
to be 20 times larger than the amplitude of the s mode. Actually, the $s$ mode couples to the 
entire continuum of modes, and hence the above model assumption is reasonable. Fig. 6 reveals 
$|E|^2$ to be 2.5 times larger in the fully nonlinear problem than when the nonlinear current 
is neglected.

Thus, we have shown that the triplet of cylindrical harmonics, better correponding to the 
curvature mechanism, is amplified more effectively than a single cylindrical harmonic. This 
means, inter alia, that the true polarization of collective curvature modes can only be 
obtained by calculating the dielectric permittivity tensor of the plasma flowing in a strong 
curved magnetic field. Here, the solution of the wave equations not only yields the dispersion 
relation for normal modes $\omega =\omega({\bf k})$ but also determines their polarization. We 
note that initially, it is totally unclear what polarization corresponds to nonstable modes. 

At first glance, the essentially nonlinear problem discussed above is not directly relevant to 
the amplification problem in the linear approximation. We included the nonlinearity only in 
order to examine the self-consistent coupling of modes $s$ and $s\pm 1$. Even in the case of 
a weak nonlinearity, the presence of adjacent modes $s\pm 1$ alters the amplification of the 
$s$ mode significantly. It is also absolutely clear that the coupling of harmonics $s\pm 1$ 
with the low-frequency harmonic $s = 1$ results in the appearance of all possible azimuthal 
harmonics.

\subsection{Calculation of the dielectric permittivity tensor}

In this section, we show that the asymptotic form of the dielectric permittivity tensor obtained 
in Ref. [2] for large values of the magnetic field curvature radius $\rho_{0}$ can be found by 
straightforward summation of responses (50) and (51) to individual cylindrical modes. We first note 
that in the case of an infinite toroidal field, there is only a response to the toroidal component 
$E_{\varphi}$  of the wave electric field [92]. Here, we only consider the case of a stationary 
medium; therefore, the time dependence can be chosen to be of the form $\exp\{- i \omega t\}$. 

Summation over all cylindrical modes yields
\begin{equation}
D_{\varphi} (\rho, \varphi) = E_{\varphi} (\rho, \varphi) -
\sum\limits_{s = - \infty}^{\infty} E_{\varphi}(\rho, s) K(\rho, s) \exp\{is\varphi\},
\label{eqm}
\end{equation}
where (see, e.g., Ref. [36])
\begin{equation}
K (\rho, s) = \frac{4 \pi e^2}{\omega} \int
\frac{v_{\varphi}}{\omega - s v_{\varphi}/\rho} \frac{\partial f^{(0)}}{\partial p_{\varphi}}
{\rm d}p_{\varphi}
\end{equation}
and $f^{(0)}(p_{\varphi})$ is the nonperturbed particle distribution function. Applying the 
Fourier transformation 
\begin{equation}
E_{\varphi}(\rho,s) = \frac{1}{2\pi} \int\limits_{0}^{2\pi} E_{\varphi}(\rho,\varphi') 
\exp\{- i s\varphi'\} {\rm d}\varphi'
\end{equation}
and passing to a Cartesian coordinate system, we find
\begin{eqnarray}
D_{x} = E_{x} + \frac{1}{2\pi}\int\frac{\rho'{\rm d}\rho'{\rm d}\varphi'}{\rho'}  
\sum\limits_{s = - \infty}^{\infty}E_{\varphi}(\rho', \varphi')\delta(\rho - \rho')
\nonumber \\
\times \, K(\rho, s) \, \exp\{i s (\varphi - \varphi')\}\sin \varphi,\\
D_{y} = E_{y} - \frac{1}{2\pi}\int\frac{\rho'{\rm d}\rho'{\rm d}\varphi'}{\rho'}
\sum\limits_{s = - \infty}^{\infty}E_{\varphi}(\rho', \varphi')\delta(\rho - \rho')
\nonumber \\
\times \, K(\rho, s) \, \exp\{i s (\varphi - \varphi')\}\cos \varphi.
\end{eqnarray}
We choose a local coordinate sytem in the particle orbit plane with the $y$ axis directed along 
the magnetic field. From the expressions presented above, we obtain the components of the kernel 
of the dielectric permittivity operator:
\begin{eqnarray}
\varepsilon_{yy}({\bf{r}}, {\bf{r'}})  =  1 - \frac{1}{2\pi}\frac{1}{\rho'}
\sum\limits_{s = - \infty}^{\infty}\delta(\rho - \rho') K(\rho, s) 
\nonumber  \\
\times \, \exp\{i s (\varphi - \varphi')\}\cos \varphi \cos \varphi';\\
\varepsilon_{yx}({\bf{r}}, {\bf{r'}})  =  \frac{1}{2\pi}\frac{1}{\rho'}
\sum\limits_{s = - \infty}^{\infty}\delta(\rho - \rho') K(\rho, s)
\nonumber  \\
\times \,\exp\{i s (\varphi - \varphi')\}\cos \varphi \sin \varphi';\\
\varepsilon_{xy}({\bf{r}}, {\bf{r'}})  =  \frac{1}{2\pi}\frac{1}{\rho'}
\sum\limits_{s = - \infty}^{\infty}\delta(\rho - \rho') K(\rho, s)
\nonumber \\
\times \,\exp\{i s (\varphi - \varphi')\}\sin \varphi \cos \varphi',\\
\varepsilon_{xx}({\bf{r}}, {\bf{r'}})  =  1 - \frac{1}{2\pi}\frac{1}{\rho'}
\sum\limits_{s = - \infty}^{\infty}\delta(\rho - \rho') K(\rho, s)
\nonumber \\
\times \,\exp\{i s (\varphi - \varphi')\}\sin \varphi \sin \varphi',
\end{eqnarray}
which determine the material relation 
\begin{equation}
D_{i}({\bf{r}}) = \int \varepsilon_{ij}({\bf{r}}, {\bf{r'}})E_{j}({\bf{r'}}) {\rm d} {\bf{r'}}.
\end{equation}

We note that the kernel found satisfies the necessary symmetry property
\begin{equation}
\varepsilon_{ij}({\bf{r}}, {\bf{r'}}, \omega) = \varepsilon_{ji} ({\bf{r'}}, {\bf{r}}, -\omega),
\end{equation}
resulting from the condition \mbox{$K(r, s, \omega) = K(r, -s, -\omega)$.} 

Further, it is well known that for calculating the dielectric permittivity tensor, only the 
symmetrized form of $\varepsilon_{ij} (\omega, {\bf {k}}, {\bf {r}})$ must be used [93, 94]: 
\begin{equation}
\varepsilon_{ij} (\omega, {\bf {k}}, {\bf {\boldsymbol\eta}} \to {\bf {r}})
= \int \varepsilon_{ij}(\omega, {\bf{\boldsymbol\xi}}, {\bf{\boldsymbol\eta}}) \exp\{-i\bf{k\boldsymbol\xi}\}{\rm d} {\bf{\xi}} .
\label{kadom}
\end{equation}
Here, by definition, ${\boldsymbol{ \boldsymbol\eta}} = ({\bf r} + {\bf r'})/2$ and  
${\bf \boldsymbol\xi} = \bf{r} - \bf{r'}$. It is important to note that only this tensor 
correctly describes the interaction between waves and particles in a medium with slowly 
varying parameters (see, e.g., Refs [35, 86]; in problems dealing with cosmological plasma, 
this approach was applied in Ref. [95]).

Substituting the components of the kernel, we now obtain
\begin{eqnarray}
\varepsilon_{xx}(\omega, {\bf {k}}, {\bf {\boldsymbol\eta}}) 
= 1 - \frac{1}{2\pi}\int {\rm d}{\bf \boldsymbol\xi} 
\exp\{-i{\bf{k \boldsymbol\xi}\}}\frac{1}{|{\bf \boldsymbol\eta 
- \boldsymbol\xi}/2|}\times\nonumber\\\sum\limits_{s = - \infty}^{\infty}
\delta(|{\bf \boldsymbol\eta + \boldsymbol\xi}/2| -|{\bf \boldsymbol\eta - \boldsymbol\xi}/2|) 
K(|{\bf \boldsymbol\eta + \boldsymbol\xi}/2|, s) \times
\nonumber\\
\exp\{i s (\varphi - \varphi')\}\sin \varphi \sin \varphi',
\label{eps1} \\
\varepsilon_{xy}(\omega, {\bf {k}}, {\bf {\boldsymbol\eta}}) 
= \frac{1}{2\pi}\int {\rm d}{\bf \boldsymbol\xi} 
\exp\{-i{\bf{k \boldsymbol\xi}\}}\frac{1}{|{\bf \boldsymbol\eta 
- \boldsymbol\xi}/2|}\times\nonumber\\\sum\limits_{s = - \infty}^{\infty}
\delta(|{\bf \boldsymbol\eta + \boldsymbol\xi}/2| -|{\bf \boldsymbol\eta - \boldsymbol\xi}/2|) 
K(|{\bf \boldsymbol\eta + \boldsymbol\xi}/2|, s)\times\nonumber \\
\exp\{i s (\varphi - \varphi')\}\sin \varphi \cos \varphi',
\label{eps12} \\
\varepsilon_{yx}(\omega, {\bf {k}}, {\bf {\boldsymbol\eta}}) 
= \frac{1}{2\pi}\int {\rm d}{\bf \boldsymbol\xi} 
\exp\{-i{\bf{k \boldsymbol\xi}\}}\frac{1}{|{\bf \boldsymbol\eta 
- \boldsymbol\xi}/2|}\times\nonumber\\\sum\limits_{s = - \infty}^{\infty}
\delta(|{\bf \boldsymbol\eta + \boldsymbol\xi}/2| -|{\bf \boldsymbol\eta - \boldsymbol\xi}/2|) 
K(|{\bf \boldsymbol\eta + \boldsymbol\xi}/2|, s)\times\nonumber \\
\exp\{i s (\varphi - \varphi')\}\cos \varphi \sin \varphi',
\label{eps21} \\
\varepsilon_{yy}(\omega, {\bf {k}}, {\bf {\boldsymbol\eta}}) 
= 1 -\frac{1}{2\pi}\int {\rm d}{\bf \boldsymbol\xi} 
\exp\{-i{\bf{k \boldsymbol\xi}\}}\frac{1}{|{\bf \boldsymbol\eta 
- \boldsymbol\xi}/2|}\times\nonumber\\\sum\limits_{s = - \infty}^{\infty}
\delta(|{\bf \boldsymbol\eta + \boldsymbol\xi}/2| -|{\bf \boldsymbol\eta - \boldsymbol\xi}/2|) 
K(|{\bf \boldsymbol\eta + \boldsymbol\xi}/2|, s) \times\nonumber \\
\exp\{i s (\varphi - \varphi')\}\cos \varphi \cos \varphi'.
\label{eps3}
\end{eqnarray}
The angles  $\varphi$ and $\varphi'$ in expressions (69)--(72) are functions of the polar 
angles $\alpha_{\eta}$ and $\alpha_{\xi}$ of the vectors $\bf{\boldsymbol\eta}$ and 
$\bf{\boldsymbol\xi}$, 
\begin{eqnarray}
\sin \varphi = \frac{|\boldsymbol\eta|\sin\alpha_{\eta} 
+ (|\boldsymbol\xi|/2)\sin\alpha_{\xi}}{|{\bf \boldsymbol\eta + \boldsymbol\xi}/2|},\\
\cos \varphi' = \frac{|\boldsymbol\eta|\cos\alpha_{\eta} 
- (|\boldsymbol\xi|/2)\cos\alpha_{\xi}}{|{\bf \boldsymbol\eta - \boldsymbol\xi}/2|}.
\end{eqnarray}
As a result, the integration in (69)--(72) reduces to integration over the components of the vector 
$\bf{\boldsymbol\xi}$ that are perpendicular to $\bf{\boldsymbol\eta}$. On the other hand, the 
the delta function in relations (69)--(72) are given by 
\begin{eqnarray}
&&\delta(...)  =  \frac{\delta(\theta - \pi/2)}{(|{\bf \boldsymbol\eta 
+ \boldsymbol\xi}/2| -|{\bf \boldsymbol\eta - \boldsymbol\xi}/2|)'_{\theta}} 
\nonumber \\
&& + \frac{\delta(\theta + \pi/2)}{(|{\bf \boldsymbol\eta + \boldsymbol\xi}/2| -|{\bf \boldsymbol\eta - \boldsymbol\xi}/2|)'_{\theta}},
\label{deltaf}
\end{eqnarray}
where $\theta$ is the angle between vectors  $\bf{\boldsymbol\eta}$ and 
${\bf{\boldsymbol\xi}}$. Therefore, the integration over the angles is 
carried out in a trivial manner. Finally, performing the transition 
${\bf {\boldsymbol\eta}} \to {\bf {r}}$ we obtain 
$\cos\alpha_{\eta} \to \cos\alpha_{r} =1$. 
In accordance with (75), we can therefore write  
$({\bf k \boldsymbol\xi}) = k_{\parallel} |\boldsymbol\xi|$, 
where $k_{\parallel}$ is the component of the wave vector directed along the magnetic field.

The property of the final result being independent of $k_{\perp}$ 
is very important. Precisely it provides the same symmetry property as in a 
homogeneous medium [42]:
\begin{equation}
\varepsilon_{ij}(- \omega, -{\bf k}, -{\bf B}, {\bf r}) = \varepsilon_{ji}(\omega, {\bf k}, {\bf B}, {\bf r}).
\end{equation}
If transformation (68) is neglected, the necessary symmetry cannot be obtained [36] (the authors 
of Ref. [36] explained the dependence of the dielectric tensor on $k_{\perp}$ by $k_{\perp}$ not 
being a Killing vector).

Finally, we use the Taylor expansion in $|\boldsymbol\xi|$ and the reduction of summation to a 
delta function:
\begin{eqnarray}
&&\sum(...)\frac{1}{\omega|{\bf \boldsymbol\eta + \boldsymbol\xi}/2|/v_{\varphi} - s}
\nonumber \\
&&\to i\pi \int (...) \delta \left[s 
- \frac{\omega(|\boldsymbol\eta|^2 + |\boldsymbol\xi|^2/4)^{1/2}}{v_{\varphi}}\right]{\rm d}s.
\end{eqnarray}
We then obtain
\begin{eqnarray}
\varepsilon_{xx}  =  1 - i\frac{8\pi^2 e^2}{\omega}\int F''(\kappa) \
frac{v_{\varphi}}{\omega}\frac{\partial f^{(0)}}{\partial p_{\varphi}}{\rm d} p_{\varphi},\\
\varepsilon_{xy}  =  - \varepsilon_{yx}   =   \frac{8\pi^2 e^2}{\omega}\int F'(\kappa) 
\frac{\rho^{1/3}_0 v^{2/3}_{\varphi}}{\omega^{2/3}}\frac{\partial f^{(0)}}{\partial p_{\varphi}}{\rm d} p_{\varphi},\\
\varepsilon_{yy}  =  1 - i\frac{8\pi^2 e^2}{\omega}\int F(\kappa) 
\frac{\rho^{2/3}_0 v^{ 1/3}_{\varphi}}{\omega^{1/3}}\frac{\partial f^{(0)}}{\partial p_{\varphi}}{\rm d} p_{\varphi},
\end{eqnarray}
where
\begin{eqnarray}
F(\kappa) & = & \frac{1}{\pi}\int\limits^{+\infty}_{0} \exp\{i\kappa t + it^3/3\} {\rm d} t,\\
\kappa & = & \frac{2(\omega - k_{\parallel} v_{\varphi})}{\omega^{1/3} v^{2/3}_{\varphi}}\rho^{2/3}_{0},
\end{eqnarray}
the derivative of $F$ with respect to its argument $\kappa$ is indicated by a prime, and  $\rho_0$ 
is the curvature radius of the magnetic field. 

It can be readily verified that the condition  $\kappa \gg 1$ is satisfied in the magnetosphere of radio pulsars 
owing to the large curvature radius $\rho_0$ of the magnetic field, and therefore we can use the asymptotic form 
of the function  $F(\kappa)$ 
\begin{equation}
F(\kappa) \approx \frac{i}{\pi \kappa} + \frac{2i}{\pi \kappa^4} + ...
\end{equation}
Integrating by parts, we obtain the final result 
\begin{equation}
\varepsilon_{ij} =
\begin{pmatrix}
1 - \frac{3}{2}\left<\frac{\omega^2_{\rm p} v^2_{\parallel}}{\gamma^3 \rho^2_0 
\tilde{\omega}^4}\right> && 
- i\left<\frac{\omega^2_{\rm p} v_{\parallel}}{\gamma^3 \rho_0 \tilde{\omega}^3}\right> \cr
 i\left<\frac{\omega^2_{\rm p} v_{\parallel}}{\gamma^3 \rho_0 \tilde{\omega}^3}
 \right>&& 1 -  \left<\frac{\omega^2_{\rm p}}{\gamma^3 \tilde{\omega}^2}\right>
 \cr
\end{pmatrix},
\label{BGItensor}
\end{equation}
where, by definition, $\tilde{\omega} = \omega - {\bf k v}$, and the angular brackets stand both 
for averaging over the particle distribution function $f_{e^{+}, e^{-}}(p_{\varphi})$ and for 
summation over the particle species:
\begin{equation}
<(...)> \, = \sum_{e^{+}e^{-}} \int (...)  f^{(0)}_{e^{+},e^{-}}(p_{\varphi}){\rm d} p_{\varphi}.
\end{equation}

We see that dielectric tensor (84) coincides exactly with the tensor obtained in 
Ref. [2] and that it is precisely the dielectric permittivity tensor that predicts the existence 
of unstable plasma-curvature modes. As expected, in the limit $\rho_{0} \rightarrow \infty$ 
the tensor found coincides with the tensor for homogeneous plasma. The nonzero components 
$\varepsilon_{xy}, \varepsilon_{yx}$ and $\delta\varepsilon_{xx}=\varepsilon_{xx}-1$ of
$\varepsilon_{ij}$ in (84) for a finite curvature are due to the nonlocal response of 
the plasma to the electromagnetic field of the wave. The nonlocality parameter
$(v_\parallel/\tilde{\omega})/\rho_0$ is the ratio of the radiation formation length  
$l_{\rm f} = c\tau$  to the curvature radius. For the vacuum, $\tilde{\omega}\simeq\omega/\gamma^2$ 
and the length  $v_\parallel/\tilde{\omega}$ coincides with the formation length of the curvature 
radiation $l_{\rm f}$.

It is important that the components $\varepsilon_{xy}=-\varepsilon_{yx}$ and
$\delta\varepsilon_{xx}$ alter the wave polarization significantly. The relation 
between the components $E_\varphi$ and $E_\rho$  of the wave electric field, 
following from expression (84) for the dielectric permittivity tensor, is given by
\begin{equation}
\left(\varepsilon_{xy}+n_\rho n_\varphi\right)E_\varphi+\left(\delta\varepsilon_{xx}
+1-n_\varphi^2\right)E_\rho=0,
\end{equation}
where $n_\rho$ and $n_\varphi$ are dimensionless components of the wave vector 
${\bf n} = {\bf k}c/\omega$. In the case of strictly azimuthal propagation (i.e., for 
$n_\rho=0$), 
\begin{equation}
E_{\varphi} \simeq 
\frac{\delta\varepsilon_{xx}}{\varepsilon_{xy}}E_\rho
\simeq \frac{c}{\rho_c\tilde{\omega}}E_\rho.
\end{equation}
As a result, the electric field of the wave can perform negative work on the 
particle current $j_\varphi$,  i.e., it can be excited. This property is violated if 
$\delta\varepsilon_{xx} = \varepsilon_{xy}=0$, whence it follows that $E_\varphi=0$.

Thus, as we have shown (also see Ref. [96]), a wave with a polarization 
\mbox{$[E_\rho(\rho); \, E_\varphi(\rho)]\exp\{is\varphi\}$,} containing 
a single azimuthal harmonic $s$, satisfies only the Cherenkov radiation
mechanism, while the emission of a charged particle in a circular magnetic 
field in the vacuum involves three azimuthal harmonics, $s$ and $s\pm 1$. 
This property provides an exit for the wave from the phase synchronism of 
the particle currrent. Further, in the case of collective curvature radiation, 
it was shown that the hydrodynamic model of plasma moving along an infinite 
magnetic field gives different results for wave amplification depending on 
the wave polarization. Consequently, there is no other way of finding the 
polarization of unstable modes except by calculating the response of the 
medium to the electromagnetic field, or, to be more precise, calculating 
the dielectric permittivity tensor. The correct procedure for calculation 
of the dielectric permittivity tensor via expansion into cylindrical modes 
is indicated above. The derived tensor coincides with the one obtained 
previously by another method [3].

In conclusion, we note that the unsuccessful attempts to find collective 
curvature radiation have led to the introduction of the term 'curvature-drift 
instability' [36]. As shown above, the polarization chosen in such an analysis, 
namely, a single azimuthal harmonic, only ensures the possibility of the Cherenkov 
amplification mechanism. In this case, the centrifugal drift of particles plays a 
decisive role. This is virtually a single curvature effect in such an analysis. 
The Cherenkov resonance, with account of the drift motion, at best provides a 
small wave amplification [37].

\subsection{Propagation theory of radio waves}

In this section, we make several comments concerning the most recent work on the theory 
of wave propagation in the magnetosphere of pulsars and on the formation problem of their 
average pulses. Over many years, a large number of polarimetric observations have been 
accumulated [97-100], and the hollow cone model, at least in its simplest realization, 
is not sufficient for their analysis. We recall that this model is based on the following 
three assumptions (see, e.g., Ref. [15]):
\begin{itemize}
\item
the formation of polarization occurs at the point of emission;
\item
radio waves propagate along straight lines;
\item
cyclotron absorption can be neglected.
\end{itemize}
But all these assumptions turned out to be incorrect. It was shown in Ref. [101] that 
in the innermost regions of the magnetoshpere, the refraction of one of the normal modes 
is significant. After publication of the work of Mikhailovskii's group [102], it became 
clear that cyclotron absorption can significantly affect the radio emission intensity. 
The influence of the magnetosphere plasma on variation of the polarization of radio 
emission propagating in the internal regions of the magnetosphere also must not be 
neglected [103]. Here, the main point is the effect of limiting polarization, which 
consists in the following. The polarization of radio emission in the region of dense 
plasma satisfies the laws of geometric optics; therefore, the orientation of the 
polarization ellipse coincides with the magnetic field orientation in the picture plane. 
But the wave polarization in the vacuum region remains unchanged. Hence, there is a 
transition layer, after passing through which the polarization is no longer affected 
by the magnetospheric plasma. In the case of typical parameters of the pulsar magnetosphere, 
it turns out that the formation of polarization occurs not at the emission point but at a 
distance of about $\approx 0.1 R_{\rm L}$ from it [104, 105]. Taking this effect into 
account should also explain the observed fraction of circular polarization of the order 
of (5--10)\%. Therefore, for a quantitative comparison of theoretical results on radio 
emission with observational data, it is necessary to have a consistent theory of radio 
wave propagation in the magnetoshpere.

At present, the theory of radio wave propagation in the magnetosphere of a pulsar can be 
considered to provide the necessary precision [106-110]. Four normal modes exist in the
magnetosphere [3, 16]. Two of them are plasma modes and two are electromagnetic, which 
are capable of departing from the magnetosphere. An extraordinary wave (the X-mode) with 
the polarization perpendicular to the magnetic field in the picture plane propagates 
along a straight line, while an ordinary wave (the O-mode) undergoes refraction and 
deviates from the magnetic axis [101]. An important point here is that for typical 
magnetosphere parameters, refraction occurs at distances up to $\approx 0.01 R_{\rm L}$,
i.e., it can be considered separately from the cyclotron absorption and the limiting 
polarization.

Based on the Kravtsov-Orlov method [109], we have developed [110] such a theory of wave 
propagation in a realistic pulsar magnetosphere taking corrections to the dipole magnetosphere 
into account (based on the results obtained by numerical simulation in Ref. [20]) together 
with the drift of plasma particles in crossed electric and magnetic fields and a realistic 
particle distribution function. The theory developed allows dealing with an arbitrary profile 
of the spatial plasma distribution, which may differ from the one in the hollow-cone model, 
because precisely the inhomogeneous plasma distribution leads to the characteristic 'patchy' 
directivity pattern [98].

The main result consists in the prediction of a correlation between the sign of the circular 
polarization (the Stokes parameter $V$) and the sign of the derivative of the change in the 
polarization of the position angle, p.a., along the profile, $p.a.$, along the profile, 
${\rm d}p.a./{\rm d}\phi$, where $\phi$ is the phase of the radio pulse. For the ordinary mode, 
these signs must be opposite to each other, while for the extraordinary mode, they must coincide. 
As was noted, refraction of the ordinary wave leads to a deviation of beams from the rotation 
axis, and therefore the ordinary wave pattern should be broader than for the extraordinary wave. 
In the case of the ordinary mode, double radio emission profiles should mainly be observed, while 
single profiles should be observed in the case of the narrower extraordinary mode [3].

Observations also fully confirm this conclusion of the theory [111]. In reviews [99, 100], to 
perform an analysis, over 70 pulsars were chosen for which both the variation of the position 
angle and the sign of the circular polarization $V$ could be traced well (the results of the 
analysis are presented in the Table). In the case of opposite signs of the derivative 
${\rm d}p.a./{\rm d} \phi$ and the Stokes parameter  $V$, the pulsar was placed 
in class O, while in the case of identical signs, it was placed in class X. As can be seen 
from the Table, most of the pulsars exhibiting a double-peaked (index D) profile indeed 
correspond to the ordinary wave, while most of the pulsars with single-peaked profiles (index 
S) correspond to the extraordinary wave. Moreover, the average width of the radiation 
pattern at a $50\%$ intensity level $W_{50}$ (normalized with account of different pulsar 
periods $P$) for ${\rm O}_{\rm D}$ pulsars is indeed about two times larger than the average 
width of the radiation pattern for ${\rm X}_{\rm S}$ pulsars. The existence of a certain number 
of pulsars of classes ${\rm O}_{\rm S}$ and ${\rm X}_{\rm D}$ should not give rise to surpise, 
because for central passage through the directivity pattern, independently of whether it 
corresponds to the O-mode or to the X-mode, a double-peaked profile should be observed, 
while for lateral passage, a single-peaked profile should be observed.

\begin{table}
\caption{Statistics of pulsars with known circular polarization $V$ and variation of position 
angle $p.a$. The pulsar periods $P$ are expressed in seconds, and the window width $ W_ {50} $ 
in degrees.}
\vspace{.3cm}
\centering
 \begin{tabular}{|c|c|c|c|c|} 
 \hline
Profile& ${\rm O_{S}}$ & ${\rm O_{D}} $ & ${\rm X_{S}} $ & ${\rm X_{D}} $\\
\hline
Number 
&6&23&45&6\\
\hline
$\sqrt{P}W_{50}$&6.8$\pm$ 3.1&10.7$\pm$ 4.5&6.5$\pm$ 2.9&5.3$\pm$ 3.0\\
\hline
\end{tabular}
\label{table1} 
\end{table}

Another important result consists in the determination of the applicability range for the standard 
relation
\begin{equation}
p.a. = {\rm arctan} \left(\frac{\sin \chi \sin \phi}
{\sin \chi \cos \zeta \cos \phi - \sin \zeta \cos \chi}\right),
\label{p.a.}
\end{equation}
describing variation of the position angle in the mean profile under the assumption of the 
validity of the hollow-cone model (the absence of any absorption and the presence of a dipole 
magnetic field in the radiation region, precisely where the polarization is determined). Here, 
once again, $\chi$ is the inclination angle of the magnetic dipole to the rotation axis, 
$\zeta$  is the angle between the rotation axis and the direction toward the observer, 
and $\phi$ is the phase of the pulse. Accurately taking propagation effects into account has 
shown [108, 111] that such a variation of the position angle can be realized only under conditions 
of low plasma density or high mean particle energy. Significant deviations from prediction (88) 
were obtained in the case of quite reasonable parameters (for example, the multiplicity
$\lambda \sim 10^4$ and the average Lorentz factor  $\gamma \sim 50$), satisfying particle 
production models. We recall that precisely equation (88) has been for many years in estimating 
the pulsar inclination angle, which is a very important parameter for determining the structure 
of the magnetosphere. In the nearest future, we plan to perform a detailed comparison of 
observational data and theoretical predictions. 

\section{Conclusion} 

In our opinion, quite a sufficient number of arguments have been presented in this article 
to assert with confidence that the theory of the magnetosphere of radio pulsars and of the 
mechanism of their coherent radiation, developed by us in the 1980s [3], contains no internal 
contradictions. Moreover, as shown above, observational data obtained recently confirm the 
validity of the main conclusions of the theory. Therefore, with the most recent results [110] 
concerning the effects of wave propagation in the magnetosphere of a neutron star (see Section 
3.5), the theory developed by us, unlike many others, permits making quantitative predictions 
concerning the evolution of neutron stars and properties of the observed radio emission.

Once again, we stress that the present article only concerns the theoretical foundation of 
our model of radio pulsars. A more detailed exposition of quantitative predictions of the 
theory and their comparison with observational data must be discussed separately. This will 
be discussed elsewhere.

The authors are grateful to A V Gurevich for the valuable comments and support and to 
A Spitkovsky and A D Tchekhovskoy for the useful discussions. This work was partly 
supported by the Russian Foundation for Basic Research (grant 11-02-01021).

\vspace{0.3cm}

\noindent
{\bf References}

\vspace{0.3cm}

{\small

\noindent
1.      Beskin V S, Gurevich A V, Istomin Ya N {\it Zh. Eksp. Teor. Fiz.} {\bf 85} 401 (1983) 
[{\it Sov. Phys. JETP} {\bf 58} 235 (1983)] \\
2.      Beskin V S, Gurevich A V, Istomin Ya N {\it Astrophys. Space Sci.} {\bf 146} 205 (1988) \\
3.      Beskin V S, Gurevich A V, Istomin Ya N {\it Physics of the Pulsar Magnetosphere} 
(Cambridge: Cambridge Univ. Press, 1993) \\
4.      Hewish A et al. {\it Nature}, {\bf 217} 709 (1968) \\
5.      Ginzburg V L, Zheleznyakov V V {\it Annu. Rev. Astron. Astrophys.} {\bf 13} 511 (1975)\\
6.      Ginzburg V L, Zheleznyakov V V, Zaitsev V V {\it Usp. Fiz. Nauk}  {\bf 98} 201 (1969) [{\it Sov. Phys. Usp.} {\bf 12} 378 (1969)]\\
7.      Kadomtsev B B, Kudryavtsev V S {\it Pis'ma zh. Eksp. Teor. Fiz.}  {\bf 13}  61 (1971) [{\it JETP Lett.} {\bf 13} 42 (1971)] \\
8.      Lominadze D G, Mikhailovskii A B, Sagdeev R Z {\it Zh. Eksp. Teor. Fiz.}  {\bf 77} 1951 (1979) [{\it Sov. Phys. JETP} {\bf 50} 927 (1979)]\\
9.      Lominadze J G, Machabeli G Z, Usov V V {\it  Astrophys. Space Sci.}  {\bf 90}  19 (1983)\\
10.     Goldreich P, Julian W H {\it Astrophys. J.}  {\bf 157} 869 (1969)\\
11.     Coppi B, Pegoraro F {\it Ann. Physics}  {\bf 119}  97 (1979)\\
12.     Melrose D  {\it Astrophys. J.}  {\bf225} 557 (1978)\\
13.     Mestel L {\it Stellar Magnetism} (New York: Oxford Univ. Press, 1999)\\
14.     Ruderman M A, Sutherland P G {\it Astrophys. J.}  {\bf 196} 51 (1975)\\
15.     Manchester R N, Taylor J H {\it Pulsars}  (San Francisco: W. H. Freeman, 1977) \\
16.     Lyne A G, Graham-Smith F {\it Pulsar Astronomy}  (Cambridge: Cambridge Univ. Press, 1998)\\
17.     Sturrock P A {\it Astrophys. J.}  {\bf 164} 529 (1971)\\
18.     Bogovalov S V {\it Astron. Astrophys.}  {\bf 349}  1017 (1999)\\
19.     Lyubarsky Y, Kirk J G {\it Astrophys. J.}  {\bf 547} 437 (2001)\\
20.     Spitkovsky A {\it Astrophys. J.}  {\bf 648} L51 (2006)\\
21.     Kalapotharakos C, Contopoulos I {\it Astron. Astrophys.}  {\bf 496} 495 (2009)\\
22.     Michel F C {\it Astrophys. J.}  {\bf 180} 207 (1973)\\
23.     Mestel L, Wang Y-M {\it Mon. Not. R. Astron. Soc.}  {\bf 188}  799 (1979)\\
24.     Contopoulos I, Kazanas D, Fendt Ch {\it Astrophys. J.}  {\bf 511}  351 (1999)\\
25.     Goodwin S P et al. {\it Mon. Not. R. Astron. Soc.}  {\bf 349} 213 (2004)\\
26.     Gruzinov A {\it Phys. Rev. Lett.}   {\bf 94} 021101 (2005)\\
27.     Komissarov S S {\it Mon. Not. R. Astron. Soc.}   {\bf 367} 19 (2006)\\
28.     McKinney J {\it Mon. Not. R. Astron. Soc.}   {\bf 368} L30 (2006)\\
29.     Timokhin A N {\it Mon. Not. R. Astron. Soc.}  {\bf 68} 1055 (2006)\\
30.     Beskin V S {\it Usp. Fiz. Nauk }  {\bf 169} 1169 (1999) [{\it Phys. Usp.} 42 1071 (1999)]\\
31.     Usov V V, in {\it On the Present and Future of Pulsar Astronomy}, 
26th Meeting of the IAU, Prague, Czech Republic (2006) \\
32.     Lyubarsky Yu {\it AIP Conf. Proc.}   {\bf 983} 29 (2008)\\
33.     Radhakrishnan V, Cooke D J {\it Astrophys. Lett.}   {\bf 3} 225 (1969)\\
34.     Beskin V S, Gurevich A V, Istomin Ya N {\it Astrophys. Space Sci.}   {\bf 102} 301(1984)\\
35.     Beskin V S, Gurevich A V, Istomin Ya N {\it Zh. Eksp. Teor. Fiz.}   {\bf 92} 1277 (1987) 
[{\it Sov. Phys. JETP} {\bf 65} 715 (1987)]\\
36.     Lyutikov M, Machabeli G, Blandford R {\it Astrophys. J.}   {\bf 512} 804 (1999)\\
37.     Kaganovich A, Lyubarsky Yu {\it Astrophys. J.} {\bf 721} 1164 (2010)\\
38.     Larroche O, Pellat R {\it Phys. Rev. Lett.}   {\bf 59} 1104 (1987)\\
39.     MelroseDB/.{\it Astrophys. Astron.}   {\bf 16} 137 (1995)\\
40.     Beskin V S, Gurevich A V, Istomin Ya N, Arons J {\it Phys. Today}   {\bf 41} (10) 71 (1994)\\
41.     Beskin V S, Gurevich A V, Istomin Ya N {\it Phys. Rev. Lett.}  {\bf 61} 649 (1988)\\
42.     Istomin Ya N {\it Plasma Phys. Control. Fusion } {\bf 36} 1081 (1994)\\
43.     Landau L D, Lifshitz E M {\it Teoriya polya} (Moscow: Nauka, 1973) {\it The Classical Theory of Fields} (Oxford: Butterworth-Heinemann, 1980)]\\
44.     Henriksen R N, Norton J A {\it Astrophys. J.}   {\bf 201} 719 (1975)\\
45.     Mestel L, Panagi P, Shibata S {\it Mon. Not. R. Astron. Soc.} {\bf 309} 388 (1999)\\
46.     Arons J {\it Astrophys. J.} {\bf 248} 1099 (1981)\\
47.     Bai X-N, Spitkovsky A {\it Astrophys. J.} {\bf 715} 1282 (2010)\\
48.     Beskin V S {\it Usp. Fiz. Nauk} {\bf 180} 1241 (2010) [{\it Phys. Usp.} {\bf 53} 1199 (2010)]\\
49.     Kramer M et al. {\it Science} {\bf 312} 549 (2006)\\
50.     Camilo F et al. {\it Astrophys. J.}  {\bf 746} 63 (2012); arXiv:l 111.5870\\
51.     Beskin V S, Nokhrina E E {\it Astophys. Space Sci.} {\bf 308} 569 (2007)\\
52.     Gurevich A V, Istomin Ya N {\it Mon. Not. R. Astron. Soc.} {\bf 377} 1663 (2007)\\
53.     Li J, Spitkovsky A, Tchekhovskoy {\it A Astrophys. J.} {\bf 746} 24 (2012)\\
54.     Istomin Ya N, Shabanova T V {\it Astron. Zh.}  {\bf 84} 139 (2007) [{\it Astron. Rep.} {\bf 51} 119 (2007)]\\
55.     Lyne A G, Manchester R N, Taylor J H {\it Mon. Not. R. Astron. Soc.} {\bf 213} 613 (1985) \\
56.     Tauris T M, Manchester R N {\it Mon. Not. R. Astron. Soc.} {\bf 298} 625 (1998)\\
57.     Young M D T  et al. {\it Mon. Not. R. Astron. Soc.} {\bf 402} 1317 (2010)\\
58.     Beskin V S, Nokhrina E E {\it Pis'ma Astron. Zh.}  {\bf 30} 754 (2004) [{\it Astron. Lett.} {\bf 30} 685 (2004)]\\
59.     Beskin V S, Eliseeva S A {\it Pis'ma Astron. Zh.}   {\bf 31} 290 (2005) [{\it Astron. Lett.} {\bf 31} 263 (2005)]\\
60.     Eliseeva S A, Popov S B, Beskin V S, astro-ph/0611320\\
61.     Lipunov V M, Postnov K A, Prokhorov M E {\it Astron. Astophys.}  {\bf 310} 489 (1996)\\
62.     Story S A, Gonthier P L, Harding A {\it Astrophys. J.}  {\bf 671} 713 (2007)\\
63.     Popov S B, Prokhorov M E {\it Usp. Fiz. Nauk} {\bf 177} 1179 (2007) [{\it Phys Usp} {\bf 50} 1123 (2007)]\\
64.     Michel F C {\it Astrophys. J.}  {\bf 180} L133 (1973)\\
65.     Mestel L {\it Astrophys. Space Sci.} {\bf 24} 289 (1973)\\
66.     Lyubarskii Y E {\it Pis'ma Astron. Zh. } {\bf 16} 34 (1990) [{\it Sov. Astron. Lett.} {\bf 16} 16 (1990)]\\
67.     Beskin V S {\it Osecimmetrichnye statsionarnye techeniya v astrofizike} ({\it Axisymmetric Flows in Astrophysics}) 
(Moscow: Fizmatlit, 2005) [Beskin V S {\it MHD Flows in Compact Astrophysical Objects} (Heidelberg: Springer, 2010)] \\
68.     Kennel C F, Coroniti F V {\it Astrophys. J.}   {\bf 283} 694 (1984)\\
69.     Barkov M V, Komissarov S S {\it Mon. Not. R. Astron. Soc.}   {\bf 385} 28 (2008) \\
70.     Tchekhovskoy A, McKinney J C, Narayan R {\it Mon. Not. R. Astron. Soc.} {\bf 388} 551 (2008) \\
71.     Narayan R, McKinney J C, Farmer A F {\it Mon. Not. R. Astron. Soc.}  {\bf 375} 548 (2007)\\
72.     Beskin V S, Malyshkin L M {\it Mon. Not. R. Astron. Soc.}   {\bf 298} 847 (1998)\\
73.     Istomin Ya N, Sobyanin D N {\it Zh. Eksp. Teor. Fiz.}   {\bf 136} 458 (2009) [{\it JETP} {\bf 109} 393 (2009)]\\
74.     Medin Z, Lai D {\it Mon. Not. R. Astron. Soc.}   {\bf 406} 1379 (2010)\\
75.     Timokhin A N Arons J {\it Mon. Not. R. Astron. Soc.}   {\bf 429} 20 (2013)\\ 
76.     Beskin V S, Rafikov R R {\it Mon. Not. R. Astron. Soc.}   {\bf 313} 433 (2000)\\
77.     Michel F C {\it Astrophys. J.}   {\bf 158} 727 (1969)\\
78.     Daugherty J K, Harding A {\it Astrophys. J.}   {\bf 252} 337 (1982)\\
79.     Takahashi T  et al. {\it Astrophys. J.}   {\bf 470} L89 (1996)\\
80.     Abdo A A et al. {\it Astrophys. J.}   {\bf 707} 1310 (2009)\\
81.     Bosch-Ramon V, Khangulyan D {\it Int. J. Mod. Phys.} 018347(2009)\\
82.     Aharonian F A, Bogovalov S V, Khangulyan D {\it Nature}   {\bf 482} 507 (2012)\\
83.     Khangulyan D  et al. {\it Astrophys. J.} {\bf 752} L17 (2012); arXiv:1107.4833  \\
84.     Berezinskii V S  et al. {\it Astrofizika kosmicheskikh luchei} (Ed. V L Ginzburg) 
        (Moscow: Nauka, 1990) [{]\it Astrophysics of Cosmic Rays} (Ed. V L Ginzburg) 
(Amsterdam: North-Holland, 1990)]  \\
85.     Bogovalov S V {\it et al. Mon. Not. R. Astron. Soc.}   {\bf 387} 63 (2008)\\
86.     Bernstein I B, Friedland L,{\it in Basic Plasma Physics} Vol. 1 (Handbook of Plasma Physics, Vol. 1, Eds A Galeev, R N Sudan) 
(Amsterdam: North-Holland, 1984) p. 367;
87.     Machabeli G Z {\it Plasma Phys. Control. Fusion}  {\bf 33} 1227 (1991)\\
88.     Machabeli G Z {\it Plasma Phys. Control. Fusion}   {\bf 37} 177 (1995)\\
89.     Nambu M {\it Plasma Phys. Control. Fusion }  {\bf 31} 143 (1989)\\
90.     Nambu M {\it Phys. Plasmas}   {\bf 3} 4325 (1996)\\
91.     Asseo E, Pellat R, Sol H {\it Astrophys. J.}   {\bf 266} 201 (1983)\\
92.     Asseo E, Pellat R, Rosado M {\it Astrophys. J.}   {\bf 239} 661 (1980)\\
93.     Kadomtsev B B {\it Kollektivnye yavleniya v plazme} 
({\it Collective phenomena in plasma}) 2nd ed. (Moscow: Nauka, 1988) 
[Kadomtsev ' ', in {\it Reviews of Plasma Physics} Vol. 22 (Ed. V D Shafranov) 
(New York: Kluwer Acad./Plenum Publ., 2001) p. 1]  \\
94.     Bornatici M, Kravtsov Yu A Plasma Phys. Control. Fusion v 42 255 (2000)\\
95.     Dodin  I Y, Fisch N J {\it Phys. Rev. D}   {\bf 82} 044044 (2010)\\
96.     Istomin Ya N, Philippov A A, Beskin V S {\it Mon. Not. R. Astron. Soc.} {\bf 422} 232 (2012) \\
97.     Rankin J M {\it Astrophys. J.}   {\bf 274} 333 (1983)\\
98.     Rankin J M {\it Astrophys. J.}   {\bf 352} 247 (1990)\\
99.     Weltevrede P, Johnston S {\it Mon. Not. R. Astron. Soc.} {\bf 391} 1210 (2008)\\
100.    Hankins T H, Rankin J M {\it Astron. J.} {\bf 139} 168 (2010)\\
101.    Barnard J J, Arons J {\it Astrophys. J.}   {\bf 302} 138 (1986)\\
102.    Mikhailovskii A B et al. {\it Pis'ma Astron. Zh.}   {\bf 8} 685 (1982) [{\it Sov. Astron. Lett.} {\bf 8} 369 (1982)]\\
103.    Petrova S A, Lyubarskii Yu E {\it Astron. Astrophys.}   {\bf 355} 1168 (2000)\\
104.    Cheng A F, Ruderman M A {\it Astrophys. J.}   {\bf 229} 348 (1979)\\
105.    Barnard J J {\it Astrophys. J.}   {\bf 303} 280 (1986)\\
106.    Petrova S A {\it Mon. Not. R. Astron. Soc.}   {\bf 368} 1764 (2006)\\
107.    Wang C, Lai D, Han J {\it Mon. Not. R. Astron. Soc.}   {\bf 403} 569 (2010)\\
108.    Beskin V S, Philippov A A, arXiv: 1101.5733\\
109.    Kravtsov Yu A, Orlov Yu I {\it Geometricheskaya optika neodnorodnykh sred} 
(Moscow: Nauka, 1980) 
[{\it Geometrical Optics of Inhomogeneous Media} (Berlin: Springer-Verlag, 1990)] \\
110.    Beskin V S, Philippov A A {\it Mon. Not. R. Astron. Soc.} {\bf 425} 814 (2012)\\
111.    Andrianov A S, Beskin V S {\it Pis'ma Astron. Zh.} {\bf 36} 260 (2010) [{\it Astron. Lett.} {\bf 36} 248 (2010)]\\
}

\end{document}